\documentclass[]{article}
\usepackage{graphicx}
\usepackage{amsfonts}
\def\ligne#1{\hbox to\hsize{#1}}
\def\leurre{\noindent\leftskip0pt\small\baselineskip 10pt}
\newtheorem{thm}{\textrm{\sc Theorem}}
\newtheorem{cor}{\textrm{\sc Corollary}}
\newtheorem{lemm}{\textrm{\sc Lemma}}
\newtheorem{fig}{\textrm{Figure}}
\newtheorem{tab}{\textrm{Table}}
\newtheorem{algo}{\textrm{\sc Algorithm}}
\def\boxempty{\hbox{\vbox{\hsize=7pt\offinterlineskip
\ligne{
\vrule height 7pt depth 0pt width 0.6pt
\vbox to 7pt{\hsize=5.8pt
\hrule height 0pt depth 0.6pt width 5.8pt
\vfill
\hrule height 0.6pt depth 0pt width 5.8pt
}\hskip-0.5pt
\vrule height 7pt depth 0pt width 0.6pt
}}
}}
\def\norm{\vert\vert}
\def\trep{\hrule height 1pt depth 1pt width \hsize}
\def\trfn{\hrule height 0.5pt depth 0.5pt width \hsize}
\newcounter{laform}
\def\numlaform{
\refstepcounter{laform}(\thelaform)
}
\author{Maurice {\sc Margenstern}}
\title{A new system of coordinates for the tilings $\{p,3\}$ and $\{p$$-$$2,4\}$.}
\begin{document}
\maketitle

\begin{abstract}
In this paper, a new way to define coordinates for the tiles
of the tilings $\{p,3\}$ and $\{p$$-$$2,4\}$ where the natural number $p$
satisfies \hbox{$p\geq 7$} is investigated.
\end{abstract}

\section{Introduction}

   In~\cite{mmbook1}, a system of coordinates for the tilings
$\{p,3\}$ and $\{p$$-$$2,4\}$ was proposed. In this paper, we revisit this question.
Several points where not enough investigated in~\cite{mmbook1}. As I needed an exact
account for a result which will appear in a forthcoming paper, I thought it useful
to provide this exact description which is of interest in itself. It is based
in a new look with respect to what was written in~\cite{mmbook1}

   We refer the reader to~\cite{mmbook1,mmEncyclo} for an introduction to an
algorithmic approach to the tessellations of the hyperbolic plane. The first 
section, introduces the leftmost approach and the one based on the preferred son
as defined in~\cite{mmbook1}. The novelty consists in comparing these
approaches which allows us to define the properties of both constructions in a simple
way. In a preliminary Section~\ref{treebij}, direct proofs are given for establishing
the bijection of a sector of the tessellations~$\{p$$-$$2,4\}$ and~$\{p,3\}$
with a tree.
Section~\ref{define}, page~\pageref{define}, 
defines both just mentioned 
approaches. The consequences are gathered and proved in Section~\ref{res},
page~\pageref{res}, giving the coordinates of the neighbours of a node in terms
of the coordinate of the node itself.

\section{The bijection with a tree}
\label{treebij}

   We remind the reader that a \textbf{sector} of the tessellations we
consider is the set of tiles whose centre is inside the angle defined by two rays
meeting at a point. In the case of the tilings \hbox{$\{p$$-$2,4$\}$}, where
\hbox{$p\geq7$}, we assume
that the rays meet at a vertex~$V$ of a tile~$T$ and that they support consecutive sides
of~$T$. The sector contains~$T$, $V$ is called its \textbf{vertex} and the
rays its \textbf{borders}.  Assume 
that $u$ is before $v$ in the sector while counter-clockwise turning around~$V$. We say 
that $u$, $v$ is the \textbf{first}, \textit{second} border respectively.  
In the case of $\{p,3\}$, we assume that the 
rays meet at a midpoint~$M$ of a side~$\sigma$ belonging two tow tiles~$T_1$ and~$T_2$. 
Fix an end~$V$ of~$\sigma$. Let 
$T_3$ be the third tile sharing~$V$ with~$T_1$ and~$T_2$. We assume that 
one of the rays, say~$u$, also passes through the mid-point of the side shared by~$T_1$ 
and~$T_3$ and that the other, say~$v$, also passes through the mid-point of the side 
shared by~$T_2$ 
and~$T_3$. We assume that the centre of~$T_3$ is contained in the angle defined by~$u$ 
and~$v$ at~$M$. The sector is defined in the same way as previously: it is the set of
tiles whose centre lies in the angle defined by
$u$ and~$v$ which are again called the borders of the sector.

   In~\cite{paperends}, a proof that a sector of the pentagrid is in bijection with
a tree can be found. The left-hand side part of Figure~\ref{bijectionp4} illustrates its
idea. The proof can be adapted to
the situation of the heptagrid as shown by Figure~\ref{bijectionp3}. In 
Subsections~\ref{bijp4} and~\ref{bijp3}, we extend these results to
the tessellations $\{p$$-$$2,4\}$ and $\{p,3\}$ respectively, where \hbox{$p\geq 7$}
in both cases. Figure~\ref{bijectionp4} illustrates the proof also in the
case of the tessellation $\{6,4\}$ while Figure~\ref{bijectionp3} does the same for
the tessellation $\{9,3\}$.

In the next lemmas, we shall use a tool which we call the \textbf{numbering} of the sides
of a tile~$T$. We fix a side~$\sigma$ of~$T$. Starting from~$\sigma$
and counter-clockwise turning around~$T$, we increasingly assign a number to the sides,
assigning~1 to~$\sigma$. We call the process a numbering which is defined once its side~1
is fixed. We also number the line which supports the side~$i$ by~$i$.
We also number the vertices as follows: the vertex $i$+1 is shared by the sides~$i$
and~$i$+1 for \hbox{$i\in\{1..p$$-$$1\}$} and vertex~$1$ is shared by the sides~$p$ 
and~1.

We remind here the elementary properties of a regular convex polygon~$P$ of the hyperbolic
plane. The interior angle~$2\alpha$ is fixed. There is a single isosceles triangle~$T_1$
whose basis angle is~$\alpha$ and its vertex angle is $\displaystyle{2\pi{}\over p}$.
Let $O$~be the vertex of~$T$. Replicating $p$$-$1 rotations around~$O$ produces
a copy~$Q$ of the polygon. Accordingly, the circle around~$O$ which passes through a
vertex of~$Q$ also passes through the other vertices of~$Q$, it is the 
\textbf{circumscribed} circle of~$Q$. From this, we get that
the diameter which passes through a vertex or through a mid-point, we get the same 
diameters when $p$ is odd, defines a reflection which leaves~$Q$ globally invariant.

\subsection{Tessellations $\{p$$-$$2,4\}$}
\label{bijp4}

Consider a tile~$T$.
The complement in the plane of the lines supporting its sides defines $2p$$-$$4$+1 
regions which are pairwise disjoint. One region is bounded: it is inside~$T$, the others
are infinite. Fix a numbering of~$T$. Call region~$i_1$ the region which is in contact 
with the side~$i$. Call region~$i_2$ the region which is in contact with the vertex shared
by the side~$i$ and the side~$i$+1 for $i<p$ and the side~$p$ and side~1. The 
regions~$i_1$, $i_2$ are called of type~1, type~2 respectively. 

Consider a region of type~1. We may assume that it is region 1$_1$. The
region is delimited by the lines~1, 2 and~$p$. A line~$i$ defines two half-planes.
Call \textbf{inside} of~$i$, \textbf{outside} of~$i$ the half-plane defined by the
line~$i$ which contains~$T$, does not contain~$T$ respectively. The region~1$_1$
is inside the lines~2 and~$p$ and it is outside the line~1. It is plain that
the other regions of type~1 are constructed in the same way. Consider the region~1$_2$
of type~2: it is the intersection of the outside of line~1 and the outside of line~2.

\begin{lemm}\label{visp4}
Consider a sector~$S$ of the tessellation $\{p$$-$$2,4\}$. 
Let~$T$ be a tile whose centre is inside~$S$. Fix a numbering of the sides of~$T$.
The lines~$i$ with \hbox{$3\leq i \leq p$$-$$3$} are non secant with line~$1$.
Say that a point $P$ is \textbf{visible} from a side~$\sigma$ of~$T$ if $P$~and the 
centre of~$T$  are on the same side of~$\sigma$. 
If $P$~is in a region of type~$1$, type~$2$, it is not visible from the side, the two sides 
respectively which
are in contact with this region and it is visible from all the other
sides. 
\end{lemm} 

\noindent
Proof. 
Consider the numbering indicated by the lemma. As side~2 is orthogonal
to both side~1 and side~3, the lines~1 and~3 are non-secant.
Consider the rays~$b_i$ issued from the centre of~$T$, meeting the side~$i$ and
supported by the bisector of the side~$i$. 
Let $B_i$ be the sector whose vertex is the centre of~$T$ which is delimited
by the rays $b_{i+1}$ and $b_{i-1}$ in this order, with \hbox{$2\leq i\leq p$$-$$2$}
and $B_1$ being delimited by $b_{p-2}$ and $b_2$ in this order. The angle between the 
rays is $\displaystyle{{4\pi}\over p}$.  As the side~$i$ is both orthogonal to $b_i$ 
and to the lines~$i$$-$1 and~$i$+1, where $i$$-$1 is replaced by $p$$-$2 when $i=1$,
these two lines are non-secant with~$b_i$. Consequently, the line~$i$ is completely 
contained in $B_i$. Now, by an angle argument, it is plain that
$B_i$ is disjoint from $B_1$ if and only if \hbox{$3\leq i\leq p$$-$$3$}. This
proves that the line~$i$ is non secant with line~$1$ for those values of~$i$.

Consider a point~$P$ which is outside~$T$. It is in a single region~$i_1$ or~$i_2$. 
We may assume that it is the region~1$_1$ or~1$_2$ by changing the side~$1$ of
the numbering.

First, assume that $P$ is in the region~$1_1$. 
For the values of~$i$ such that \hbox{$3\leq i\leq p$$-$$2$}, as the considered 
sectors defined by $b_i$
and $b_{i+2}$ are disjoint from that defined by $b_p$ and $b_2$, the inside of the 
line~$i$ also contains~$P$.  For the sides~$2$ and $p$$-$2, by definition of the types of 
the regions, $P$ is also in the inside of the lines both for line~1 and for 
the line~$p$$-$2. 

Secondly, assume that $P$ is in the region~$1_2$. Then $P$ is in the outside of line~$2$
and still in the inside of the line~$p$$-$2 and in the inside of the lines~$i$
for \hbox{$3\leq i\leq p$$-$$3$}.  This proves the lemma. 
\hfill\boxempty

From the lemma, we can prove another property~:

\begin{lemm}\label{projp4}
Let $S$ be a sector of the tessellation~$\{p$$-$$2,4\}$.
Let ~$T$ be a tile whose centre is in~$S$. Let $P$ be a point outside~$T$
such that its orthogonal projection on the line~$1$ of~$T$ falls inside the side~$1$
of~$T$. The orthogonal projection of~$P$ on the lines~$i$ with 
\hbox{$i\in\{3..p$$-$$3\}$} also falls inside the side~$i$. 
\end{lemm}

\noindent
Proof of Lemma~\ref{projp4}. Indeed, let ~$H$ be the orthogonal projection of~$P$ on the 
line~1 of~$T$. From the hypothesis, $H$ is inside the side~1 of~$T$.
Let~$K$ be the projection of~$P$ on the line~$i$ with \hbox{$i\in\{3..p$$-$$3\}$}.
The line $PK$ does not meet the line~$i\ominus1$: otherwise, let~$L$ be
the intersection with the line~$i\ominus1$. As the sides~$i$
and $i\ominus1$ are perpendicular, from the intersection, there would be two distinct
perpendiculars to the line~$i$ unless $P$ lies on the side~$i\ominus1$ and $H$ belongs
to the line~$PK$. But we assumed that $H$ is inside the side~1 of~$T$. A similar argument
with the line~$i\oplus1$ which is also orthogonal to the line~$i$ shows us that $K$
is inside the side~$i$. This proves the lemma. \hfill\boxempty

Note that if $P$ has its orthogonal projection on the line~1 falls inside the side~1
of~$T$, its orthogonal projection on the line~2 cannot fall inside the side~2 of~$T$
as the line~2 is orthogonal to the line~1 of~$T$. The same remark also holds for the 
line~$p$$-$2.

We are now in the position to prove the following result:

\begin{thm}\label{treebijp4}
The set of tiles of a sector~$S$ of the tessellation $\{p$$-$$2,4\}$ is in bijection
with a tree possessing to kinds of nodes, $B$ and $W$, generated by the
following two rules: \hbox{$W\rightarrow BW^{p-5}$} and \hbox{$B\rightarrow BW^{p-6}$},
the root being a $W$-node.
\end{thm}

\noindent
Proof of Theorem~\ref{treebijp4}.
Let $\cal H$ be the tile of~$S$ which has the vertex of~$S$ among its vertices.
Call it the \textbf{head} of~$S$. Let $O$ be the centre of~$\cal H$, and let be $u$ 
and~$v$ the rays defining~$S$, the first and the second borders of~$\cal H$ respectively. 
We define the \textbf{cornucopia} of the sector 
as the set of tiles in the sector which share a side with the first border. 

We attach the root of the tree to~$\cal H$. Number the sides of~$\cal H$ with its side 
on~$u$ as its side~1. Let $T_j$, with 
\hbox{$j\in\{2..p$$-$$3\}$}, be the reflection of~$\cal H$ in its side~$j$. Let $H_j$ 
be the orthogonal projection of~$O$ on the line~$j$ of~$\cal H$. 
As $\cal H$ is a regular convex polygon, $H_j$ is the midpoint of the 
side~$j$ of $\cal H$. 
Define $S_3$ as the image of~$S$ by the shift along the side~2 of~$\cal H$. For
$k\in\{3..p-4\}$, define $S_{k+1}$ as the image of~$S_k$ under the rotation around~$O$
which transforms the side~$k$ into the side~$k$+1.
Now, the complement of $\cal H$ in~$S$ can be decomposed into the $p$$-$5
sectors $S_k$ with \hbox{$k\in\{3..p$$-$$3\}$} together with a remaining region which we 
call a \textbf{strip}, denote it by~$B$. 
The head of~$S_k$ is $T_k$ as can easily be seen.
Each $T_k$ is numbered with, as its side~1, the
side~$k$ of $\cal H$. As each $S_k$ with \hbox{$k\in\{3..p$$-$$3\}$} is in the
outside of the line~$k$ of~$\cal H$, the distance from~$O$ to $S_k$ is at least
that from~$O$ to the line~$k$, so that it is $OH_k$, as $H_k$ is the orthogonal
projection of~$O$ on the line~$k$.
Denote by $C_k$, with $k\geq 1$, the tiles of the cornucopia, with $C_1 = \cal H$,
$C_2$ being its image by reflection in the side~2 of~$C_1$.
Fix the side~1 of~$C_2$ to be the side~2 of~$\cal H$. Then, for 
\hbox{$k\in\{2..p$$-$$4\}$} $C_{k+1}$ is the image of~$C_k$ by reflection on its 
side~3, the side~1 of~$C_{k+1}$ being the side~3 of~$C_k$. We note that $C_2$ is in~$B$: 
we call it the head of~$B$ which is delimited by the side~1 of~$C_2$, 
the ray~$u$ supporting the side~1 of~$C_1$, and the first border of $S_3$, which supports
the side~$p$$-$2 of~$C_2$.

Denote by $S_u$ the image of~$S$ by the shift along~$u$ which also transforms $C_1$ 
into~$C_2$. Now, if we repeat to $S_u$ the process which was performed in $S$ to define
the $S_k$'s starting from $S$, we obtain a sequence of sectors $Q_j$ with
\hbox{$j\in\{3..p$$-$$4\}$} where $Q_3$ is the image of~$S_u$ by the shift along the
side~3 of $C_2$ and then $Q_{j+1}$ is the the image of~$Q_j$ by the 
rotation around the centre of~$C_2$ which transforms its side~$j$ into its side~$j$+1.
For the numbering of $Q_j$, we fix its side~1 as 
the side~$j$ of $C_2$. 
The complement of~$C_2$ in the strip $B$ 
consists of the $Q_j$'s for \hbox{$j\in\{3..p$$-$$4\}$} and a new strip
which is the image of~$B$ under the shift which transforms $S$ into~$S_u$. As 
$B\subset S_u$, the distance from $O$ to~$B$ is at least that from $O$ to~$S_u$
which, clearly, is $OH_2$ which is also the distance from~$O$ to~$C_2$.
We say that $C_2$ is the $B$~son of $\cal H$ and that the $T_k$'s with
\hbox{$k\in\{3..p$$-$$3\}$} are its $W$-sons. The sons of~$\cal H$ constitute 
the level~1 of the tree. The sons of~$C_2$ are $C_3$, the $B$-son, and the heads of
the $Q_j$'s with \hbox{$j\in\{3..p$$-$$4\}$}, the $W$-sons.

From now on, say that the head of a strip is a $B$-node, and that the head of a sector
is a $W$-node. Define their \textbf{sons} as their image by reflection
in the sides~$j$ with \hbox{$j\in\{2..p$$-$$3\}$} for $W$-nodes and
in both \hbox{$j\in\{3..p$$-$$3\}$} for $B$-nodes. The reflection in the side~2, side~3
yields the $B$-node for $W$-, $B$-nodes, respectively.
Recursively repeating this decomposition
can be represented by a recursive application of the rules given in the theorem
which provides us with the tree stated in the theorem. This defines an injection
of the tree into~$S$.
\vskip 10pt
\vtop{
\ligne{\hfill
\includegraphics[scale=0.45]{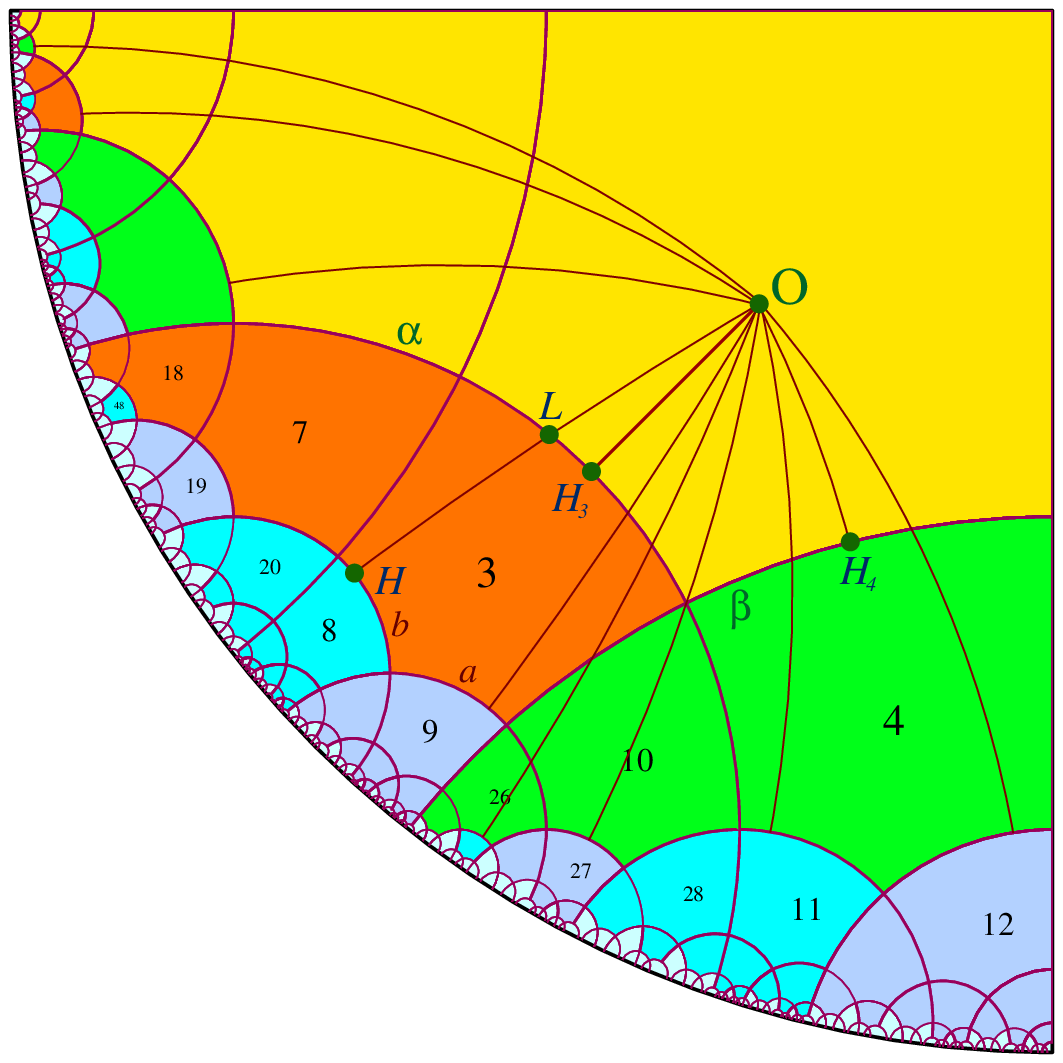}
\includegraphics[scale=0.875]{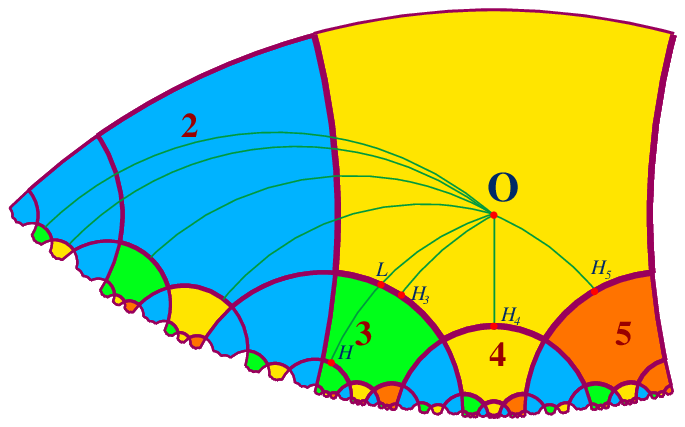}
\hfill}
\ligne{\hfill
\vtop{\leftskip 0pt\parindent 0pt
\vspace{-10pt}
\begin{fig}\label{bijectionp4}
\leurre
Proof of the bijection for the pentagrid, to left, and for the tessellation $\{6,4\}$,
to right.
It can be generalized to the tessellations $\{p$$-$$2,4\}$.
\end{fig}
}
\hfill}
}
\vskip 10pt

We proved that the distance from~$O$ to~$B$ is $OH_2$ and that the distance
from~$O$ to~$S_j$ for \hbox{$j\in\{3..p$$-$$3\}$} is $OH_j$. As $O$ is the centre
of~$\cal H$, all these distances are equal, denote by $a$ their common value.
From the previous construction, it is plain that the tree is injectively
mapped into the sector by the correspondence which associates a tile to a node.

We say that $B$ and the $S_j$'s we constructed from~$\cal H$ define
the first generation. Applying the just above defined construction to the
generation~$n$, the sons of the nodes of the generation~$n$ constitute the
generation~$n$+1.
Note that, in our construction, $O$ is visible from the side~1 of the white sons 
of~$\cal H$. Consider a white node~$T$ of the generation~$n$ and assume that $O$~belongs
to the region~$R$ of type~1 associated to the side~1 of~$T$, say~$\sigma_0$. 
From Lemma~\ref{visp4}, we obtain that $O$ is visible from the sides~1 of the $W$-sons 
of~$T$. Accordingly, Lemma~\ref{projp4} says that the orthogonal projection of~$O$ on the side~1 of a $W$-son of~$T$ lies inside that side. Fix a $W$-son of~$T$ and
let $\sigma_1$ be its side~1. Denote by~$H_0$, $H_1$ the orthogonal projection of $O$ 
on~$\sigma_0$, $\sigma_1$ respectively. Lemma~\ref{projp4} says that $H_0$, $H_1$
is inside $\sigma_0$, $\sigma_1$ respectively, see Figure~\ref{bijectionp4}. 
As $H_1$ and~$O$ are not on the same side
of the line~$\ell$ which supports~$\sigma_0$, $OH_1$ cuts~$\ell$ at~$L$. Clearly,
\hbox{$OL\geq OH_0$} as $OH_0$ is the distance from~$O$ to~$\ell_0$. Let $\ell_1$
be the line supporting $\sigma_1$. From Lemma~\ref{visp4}, $\ell_1$ and $\ell_0$
are non-secant. Accordingly, there is a point~$P_0$ on~$\ell_0$ and a point~$P_1$
on~$\ell_1$ such that $\ell_0\perp P_0P_1$ and $\ell_1\perp P_0P_1$. Hence, 
$P_0P_1$ realizes the distance between~$\ell_0$ and~$\ell_1$, so that 
$H_0L\geq P_0P_1$. If $b$ is the smallest value between $a$ and the distances $d_j$
between the line~1 of~$T$ and the lines~$j$ of~$T$ with \hbox{$j\in\{3..p$$-$$3\}$}
which are non secant with the line~1 from Lemma~\ref{visp4}, we get
that $OH_1\geq OH_0+b$. Accordingly, if we assumed that $OH_0\geq nb$,
we get that $OH_1\geq (n$+$1)b$. 
Now, the same arguments can be repeated with the sons of~$C_2$ as $O$ is visible
from its sides~$j$ with \hbox{$j\in\{3..p$$-$$3\}$}.

To get the bijection property, it is enough to prove that any point in
the angle defined by the vertex of~$S$ and its borders and by the fact it 
contains $\cal H$, belongs to at least one tile~$T$ of~$S$. Let $P$ be such a point.
If $P$ belongs to the cornucopia, as the distance from $C_k$ to $O$ is at least 
$(k$$-$$1)a$, it falls in at least on of the $C_k$'s. Note that the $C_k$'s
are successive $B$-sons of $B$-sons of~$C_1$. If not, from the above construction,
$P$ belongs to one of the~$S_j$'s with \hbox{$j\in\{3..p$$-$$3\}$} headed by
the $W$-sons of the level~1 of the tree, or to the $W$-sons of~$C_k$ with $k\geq 2$
which belong to the level~$k$. 
Note that from what we proved, the strip whose head is on the level $n$
is at a distance at least $nb$ from~$O$. Let $S^1$ be the sector in which $P$ lies. 
We repeat the same argument. If $P$ is in the cornucopia~$K_1$ of~$S^1$, we shall find 
it in a tile of~$K_1$, otherwise it will be in a sector~$S^2$ obtained from the 
process we just described. By our arguments, the distance from $O$ to~$S^1$, $S^2$ is 
at least $n_1b$, $n_2b$ for some positive integer $n_1$, $n_2$ respectively and, our 
argument proves that $n_2>n_1$. We can construct a sequence $S^1$, $S^2$, $\ldots$, $S^m$ 
containing $P$. From this observation, as $OP$ is finite, there is an $m$ such that $P$ 
is contained in the cornucopia of~$S^m$. Consequently, we shall find a tile of~$S$ 
containing~$P$ and this tile, by our above observation, will be in correspondence with 
a node of the tree.
This completes our proof. \hfill\boxempty

From the bijection property, we shall say indifferently tile or node for a tile.
When we shall use the term node, we shall make it precise whether it is a $B$- or a 
$W$-node if it is needed. From the proof of the theorem, we conclude the following property:

\begin{cor}\label{BWp4}
The $B$-nodes of the tree are 
in the cornucopias. The head of the cornucopias are the $W$-nodes exactly.
\end{cor}

\subsection{Tessellations $\{p,3\}$}\label{bijp3}

Consider a tile~$T$ of the tessellation $\{p,3\}$ with \hbox{$p\geq7$}.
We number the vertices of~$T$ such that  the vertices $i$ and $i$+1 are the ends
of the side~$i$ when \hbox{$i<p$} and the vertices~$p$ and~1 are the ends of the
side~$p$, see Figure~\ref{pseudo93} where the numbers are written only
for vertex~1.

Consider the mid-point line~1 which, by definition, joins the midpoints of
the sides~$p$ and~1. This name is grounded on the following considerations.
Let $M_i$ the be the mid-point of the side~$i$ and denote by $V_i$ the vertex~$i$.  
Let $T_1$ be the reflection of~$T$ in its side~1. Let $N_2$ be the mid-point
of the side of~$T_2$ which abuts the vertex~2 of~$T$. Then the triangles 
$N_2V_2M_1$ and $M_1V1M_p$ are isosceles triangles as the interior angles at~$V_2$ in~$T_2$ 
and~$V_1$ in~$T$ are equal. Then, the basis angles of these triangles are equal and
as $N_2$ and $M_2$ are not on the same side of line~1 as far as $T_2$ is the reflection
of~$T$ in this line, this means that the points $N_2$, $M_1$ and~$M_p$ lie on the
same line which we call mid-point line~1. Also note that the considered triangles 
being isosceles, the basis angle is acute. 
\vskip 10pt
\vtop{
\ligne{\hfill
\includegraphics[scale=1]{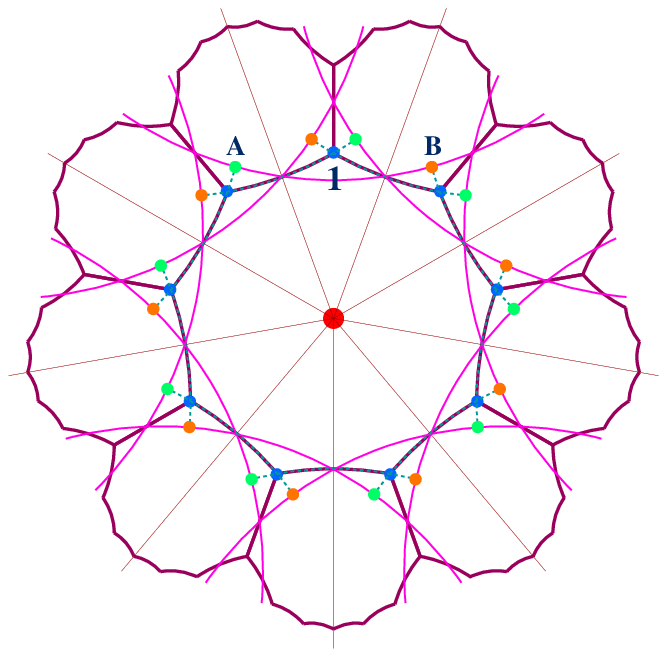}
\hfill}
\ligne{\hfill
\vtop{\leftskip 0pt\parindent 0pt
\vspace{-10pt}
\begin{fig}\label{pseudo93}
\leurre
The pseudo-sides in a tile of the tessellation $\{p,3\}$. 
\end{fig}
}
\hfill}
\vskip 5pt
}

Let~$O$ be the centre of~$T$. Note that $OM_1$ is a symmetry axis of~$T$ which leaves 
the side~1 globally invariant, so that $OM_1\perp V_1V_2$. As the basis angle 
\hbox{$(M_1V_1,M_1M_p)$} is acute, $V_1$ and~$O$ are not on the same side of~$M_1M_p$. 
Note that $OV_1$ is also a symmetry axis of~$T$ which leaves vertex~1 invariant. The 
symmetry exchanges $V_2$ and $V_p$ so that it also exchanges $M_1$ and~$M_p$. 
Consequently, \hbox{$M_1M_p\perp OV_1$}. We say that the \textbf{inside} of
the mid-point line~1 is its half-plane which contains the centre of~$T$.
At last,
note that line~2 cuts mid-point line~1 at a point~$A$ and the line~$p$$-$2
cuts mid-point line~1 at~$B$, in both cases at a right-angle. We call $AB$ the
\textbf{pseudo-side}~1 of~$T$ which, consequently, is supported by
mid-point line~1. From our last observation, the mid-point lines~$i$
and~$i$+2 with \hbox{$1\leq i<p$$-$1} are non secant. The mid-point lines~1 and~$p$$-$2
are also non-secant. Also note that the reflection in $OM_2$ exchanges the mid-point 
lines~1 and~3. As side~2 is also perpendicular to~$OM_2$, that line is non-secant
with both mid-point lines~1 and~3. From these considerations, we also get 
that the mid-point line~1 is completely included in the the angle
\hbox{$(OM_2,OM_{p-1})$}. 

By the symmetries of~$T$ which is invariant 
under any rotation around~$O$ which transforms a side of~$T$ into a side of~$T$,
these just mentioned properties can be transported to any vertex of~$T$ and
to any mid-point line which are numbered as just mentioned. 
In order to facilitate the notations, we introduce two operations:

{\leftskip 40pt\parindent 0pt \rightskip 40pt
$i\oplus j$ is $i$+$j$ if $i$+$j\leq p$, otherwise it is $i$+$j$$-$$p$;\vskip 0pt
$i\ominus j$ is $-i$$-$$j$ if $i$$-$$j\geq 1$, otherwise it is $i$$-$$j$+$p$
\par}

We are now in the position
to prove for the tessellation two lemmas which are analogous to Lemmas~\ref{visp4}
and~\ref{projp4}. We have a different notion of region as in the
case of the tessellations~$\{p$$-$2,4$\}$. Around~$T$, we consider the regions 
$R_i$ which are the intersections of the outside of the pseudo-side~$i$ with
the inside of the lines~$i\ominus2$ and~$i\oplus1$. Define $\mu_i$ to be the bisector of
the side~$i$. Then the mid-point line~$i$ and $\mu_{i\oplus1}$ are both perpendicular
to line~$i\oplus1$ so that, $\mu_{i\oplus1}$ is non-secant with the mid-point line~$i$.
Similarly, $\mu_{i\ominus2}$ is non-secant with the mid-point line~$i$ as
both those lines are perpendicular to line~$i$. Accordingly, the outside
of the mid-point line~$i$ is contained in the angle at~$O$ defined by 
$(\mu_{i\ominus2},\mu_{i\oplus1})$ whose measure is $\displaystyle{{6\pi}\over p}$.
For $p\geq 7$, this angle is less than~$\pi$. 

Let $U$ be a vertex of a tile of the tessellation~$\{p,3\}$. Three sides $s_1$, $s_2$
and~$s_3$ meet at~$U$, pairwise belonging to the three tiles~$T_1$, $T_2$ and~$T_3$ 
which meet at~$V$. Assume that $s_i$ is shared by~$T_j$ and~$T_k$ where
\hbox{$\{i,j,k\}=\{1,2,3\}$}. We may assume that the order $s_1$, $s_2$ , $s_3$
define the counter-clockwise orientation. It is then the same for the tiles $T_i$.  
Let $V$ be the mid-point of~$s_1$. Let $u$, $v$~be the ray issued from~$V$ which is 
supported by the mid-point line passing through~$V$ and through the mid-point
of~$s_2$, $s_3$ respectively. Then we define the \textbf{sector} $S$ of the 
tessellation $\{p,3\}$ defined by $V$, $u$ and~$v$ as the set of tiles
whose centers are inside the angle $(u,v)$. Note that $T_1$ belongs to~$S$.
We say that $u$ is the \textbf{first} border of~$S$, that $v$ is its \textbf{second}
one.

\ifnum 1=0 {
\begin{lemm}\label{visp3}
Consider a sector~$S$ of the tessellation $\{p,3\}$.
Let~$T$ be a tile whose centre is inside~$S$. Fix a numbering of the sides of~$T$.
The pseudo-lines~$i$ with \hbox{$i\in\{1..p\}$} are non secant with 
the pseudo-line~$j$ with \hbox{$j\not\in\{i\ominus1..i\oplus1\}$}.
Say that a point $P$ is \textbf{visible} from a pseudo-side~$\sigma$ of~$T$ if $P$~and 
the centre of~$T$  are on the same side of the pseudo-side~$\sigma$.
If $P$~is in the region~$R_i$, it may be not visible from the pseudo-sides~$j$
with $j\in\{i\ominus1..i\oplus2\}$ the two sides
respectively which
are in contact with this region and it is visible from all the other
sides.
\end{lemm}
} \fi

\vskip 10pt
\vtop{
\ligne{\hskip 10pt
\includegraphics[scale=0.95]{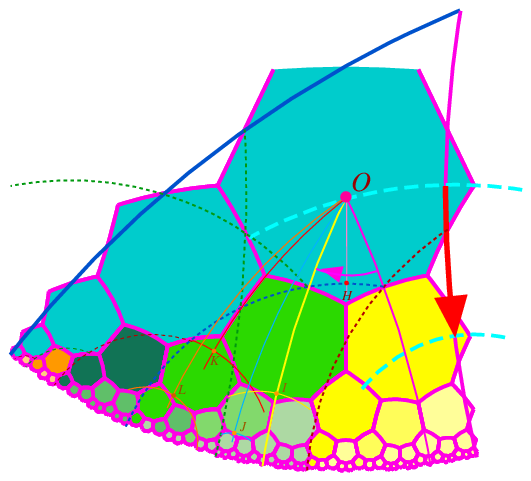}
\hskip-20pt
\includegraphics[scale=0.95]{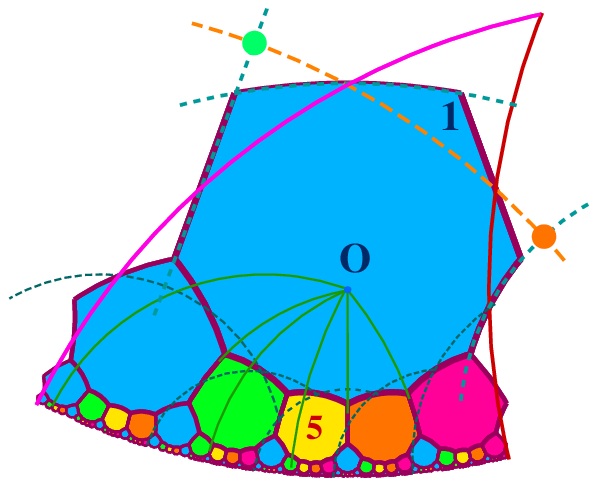}
\hfill}
\ligne{\hfill
\vtop{\leftskip 0pt\parindent 0pt
\vspace{-10pt}
\begin{fig}\label{bijectionp3}
\leurre
Proof of the bijection for the tilings $\{p,3\}$. To left, the heptagrid,
to right, an illustration for the general case with the tessellation $\{9,3\}$.
\end{fig}
}
\hfill}
\vskip 5pt
}

\begin{lemm}\label{pseudop3}
Let ~$T$ be a tile of the tessellation~$\{p,3\}$.
The mid-point lines~$i$ of~$T$ with \hbox{$i\in\{4..p-2\}$} 
are non secant with the mid-point line~$1$.
\end{lemm}

\noindent
Proof of Lemma~\ref{pseudop3}.
This comes from the above remark about the fact that the mid-point line~$i$ is completely
included in the angle defined by $(\mu_{i\ominus2},\mu_{i\oplus1})$, with
the additional property that that mid-point line is non secant with both
$\mu_{i\ominus2}$ and~$\mu_{i\oplus1}$. \hfill\boxempty

We are now in the position to prove the following result,

\begin{thm}\label{treebijp3}
The set of tiles of a sector~$S$ of the tessellation $\{p,3\}$ is in bijection
with a tree possessing to kinds of nodes, $B$ and $W$, generated by the
following two rules: \hbox{$W\rightarrow BW^{p-5}$} and \hbox{$B\rightarrow BW^{p-6}$},
the root being a $W$-node. It is the same tree as the tree defined in 
Theorem~{\rm\ref{treebijp4}} for the tessellation $\{p$$-$$2,4\}$.
\end{thm}

\noindent
Proof of Theorem~\ref{treebijp3}.
We follow the same line as in the case of Theorem~\ref{treebijp4}.
Figure~\ref{bijectionp3} illustrates the principle we apply, to the left-hand side of the 
figure, in the heptagrid, to its right-hand side, to the tessellation $\{9,3\}$.
Let $u$, $v$ be the first, second respectively border of~$S$ and let~$V$ be its vertex.
Consider $\cal H$ the head of~$S$. Let vertex~1 of~$T$ be its closest
vertex to~$V$. Denote by~$T_i$ with \hbox{$i\in\{3..p$$-$$2\}$} the reflection of~$\cal H$
in its side~$i$. Fix the side~$1$ of~$T_i$ as the side~$i$ of~$\cal H$. Then the vertex~1
of~$T_i$ is the vertex~$i\oplus1$ of~$\cal H$. Then, the shift along~$v$ which transforms 
the vertex~1 of~$\cal H$ into its vertex~$p$$-$1 also transforms~$\cal H$ into~$T_{p-2}$.
It also transforms~$S$ into a sector~$S_{p-2}$ whose vertex is the mid-point of the
side~$p$$-$1, whose first border is supported by the mid-point line~$p$$-$1 of~$\cal H$
and its second one is~$v$. Now, the successive rotations around~$O$, the centre
of~$\cal H$, transform~$S_{p-2}$ into $S_{p-j}$ with \hbox{$j\in\{3..p$$-$$3\}$}. 
Considering the complement in~$S$ of~$\cal H$ and of the $S_i$ with 
\hbox{$i\in\{4..p$$-$$2\}$}, we get a region which we again denote by~$B$ and which
we call again a \textbf{strip}. Note that $B\subset S_3$, the inclusion being proper. 

The strip~$B$, is delimited by~$u$, by the pseudo-side~4 and the mid-point line~5.
We call~$T_2$ the \textbf{head} of~$B$. The shift along the mid-point line~5 of~$\cal H$
transforms the vertex~1 of~$T_2$ into its vertex~$p$$-$1, hence the sector~$S_3$ 
into a sector~$Q_{p-2}$. Similarly, successive rotations around the centre of~$T_2$
transform $Q_{p-2}$ into $Q_{p-j}$ with \hbox{$j\in\{3..p$$-$$4\}$}. 
The head $Y_i$ of~$Q_i$ is obtained by the reflection of~$T_2$ in its side~$i$
when \hbox{$i\in\{4..p$$-$$2\}$}. It can be noted that the complement of~$T_2$
in~$B$ and of the $Q_i$'s with \hbox{$i\in\{5..p$$-$$2\}$} is again a strip,
the image of~$B$ in the shift along~$u$ which transforms~$\cal H$ into~$T_2$.
Here too, we fix that the side~1 of~$Y_i$ for \hbox{$i\in\{4..p$$-$$2\}$} 
as the side~$i$ of~$T_2$. Now, we can repeat
the argument of Theorem~\ref{treebijp4}: recursively repeating this construction
with sectors and strips, we obtain an injection from the tree defined
in the statement of Theorem~\ref{treebijp4} into the tiles of~$S$. Again, we
have to prove that the mapping is surjective.

We proceed as in the proof of Theorem~\ref{treebijp4}, using the mid-point lines
in order to estimate the distance from~$O$ to a region defined by the above
process. We note that, from the above discussion leading to Lemma~\ref{pseudop3},
for each~$i$, the vertex~$i$ of a tile~$T$ is not in the same side as the centre of~$T$
with respect to the mid-point line~$i$. The proof of the mid-point line property
show us that for any~$i$, the centres of~$T_i$ and $T_{i\oplus1}$ are outside
the mid-point line~$i$ of~$\cal H$. The distance from~$O$ to these centres
can be estimated by the distance from~$O$ to the mid-point line~$i$ of~$T_i$.
Accordingly, this distance estimates the distance from~$O$ to both $S_i$ 
and~$S_{i\oplus1}$. Note that the mid-point line~3 of~$\cal H$ allows us to estimate
the distance from~$O$ to~$B$.

As previously, we define the \textbf{cornucopia} of~$S$ as the set of its tiles
which share a side with~$u$ and $\cal H$ is called the head of the cornucopia. 
The complement of the cornucopia in~$S$ can be split
as a union of sectors which, sectors are set of tiles, are pairwise disjoint.
This construction can be repeated for all sectors.

Presently, assume that the distance from~$O$ to any sector or strip whose
head belongs to the level~$n$ of the tree is~$n.a$, where~$a$ is the smallest distance
from mid-point line~1 to the mid-point lines~$i$ of the same tile which are non secant
with it. Let $T$ be the head of a sector or a cornucopia which is a node of the 
level~$n$+1.
Let $H$ be the orthogonal projection of~$O$ in the mid-point line~1 or~2 of~$T$,
depending on the position of~$T$ with respect to its father $U$ in the tree,
call~$\ell$ this line. If $U$ is 
the head of a sector~$S_0$ or of a cornucopia,
its distance from~$O$ is measured by the orthogonal projection of~$O$ on the mid-point
line~1 or~2 of~$U$, denote it by~$\mu$. Now, from the construction, the centre
of~$T$ is not on the same side of~$\mu$ as~$O$. Accordingly, $OH$ cuts~$\mu$ at~$L$.
Now, by induction, as $U$ belongs to the level~$n$, $OL\geq na$. Also,
as $\ell$ and~$\mu$ are non-secant, $LH\geq a$. Accordingly \hbox{$OH\geq (n$+$1)a$},
which completes the proof of the theorem. \hfill\boxempty

%
\section{The leftmost and the preferred son approaches}
\label{define}


   From Theorems~\ref{treebijp4} and~\ref{treebijp3}, we know that a sector
of the tessellation $\{p$$-$$2,4\}$ or one of the tessellation $\{p,3\}$ with
the same value of~$p$, $p\geq7$, are both in bijection with the same tree.
Now, taking into consideration that the tree is embedded in the dual
graph of the tilings, we shall see that we can define infinitely many trees
which are in bijection with any sector of those tessellations. Subsections~\ref{leftmost},
and~\ref{preferred} define these trees. Subsection~\ref{leftmost} is
based on the construction performed in Subsection~\ref{preferred}.
But the definition of one the trees relies
on a notion we take from the property stated in Theorems~\ref{treebijp4}
and~\ref{treebijp3}. We turn to this point in Subsection~\ref{metal}. 

\subsection{The coordinates of the nodes}
\label{metal}

   In both Theorems~\ref{treebijp4} and~\ref{treebijp3}, the tree has two kinds of nodes,
$B$-nodes and $W$-nodes, and it is constructed by the recursive application of the 
following rules:
\vskip 5pt
\ligne{\hfill
$\vcenter{
\vtop{\leftskip 0pt\parindent 0pt \hsize=85pt
$\hbox to 15pt{$B$\hfill}\rightarrow\hskip 8pt BW^{p-6}$\vskip 0pt
$\hbox to 15pt{$W$\hfill}\rightarrow\hskip 8pt BW^{p-5}$
}
}$
\hfill\numlaform\hskip 20pt}
\newcounter{matWB}
\setcounter{matWB}{0}
\addtocounter{matWB}{\value{laform}}
\vskip 5pt
Note that we can also associate to~(\thematWB) the matrix
$\vcenter{
\vtop{\leftskip 0pt\parindent 0pt\hsize 32.5pt
\ligne{\hbox to 22.5pt{$p$$-$6\hfill} 1\hfill}
\ligne{\hbox to 22.5pt{$p$$-$5\hfill} 1\hfill}}
}$. 
From this, if $w_n$, $v_n$ is the number of $W$-nodes, $B$-nodes respectively on the 
level~$n$ of the tree, we get that:
\vskip 5pt
\ligne{\hfill
$\vcenter{\vtop{\leftskip 0pt\parindent 0pt \hsize=130pt
$v_0=0$, $w_0=1$\vskip 0pt
$\hbox to 20pt{$v_{n+1}$\hfill}=v_n + w_n$\vskip 0pt
$\hbox to 20pt{$w_{n+1}$\hfill}=(p$$-$$6)v_n + (p$$-$$5)w_n$
}}$
\hfill\numlaform\hskip 20pt}
\newcounter{recWB}
\setcounter{recWB}{0}
\addtocounter{recWB}{\value{laform}}
\vskip 5pt
Denote by $u_n$ the number of nodes on the level~$n$. As each node produces one $B$-son
exactly, $v_{n+1}=u_n$, so that, summing the last two lines of~(\therecWB), we obtain:
\vskip 5pt
\ligne{\hfill
$\vcenter{
\vtop{\leftskip 0pt\parindent 0pt \hsize=120pt
$u_{-1} = 0$, $u_0=1$\vskip 0pt
$u_{n+2} = (p$$-$$4)u_{n+1}-u_n$
}}$
\hfill\numlaform\hskip 20pt}
\newcounter{recseq}
\setcounter{recseq}{0}
\addtocounter{recseq}{\value{laform}}
\vskip 5pt
From that, we derive a matrix: 
$\vcenter{
\vtop{\leftskip 0pt\parindent 0pt\hsize 35pt
\ligne{\hbox to 22.5pt{$p$$-$4\hfill} $-1$\hfill}
\ligne{\hbox to 22.5pt{\hskip 5pt1\hfill} \hskip 5pt0\hfill}}
}$ 
whose characteristic polynomial is 
\hbox{$P(X)=X^2-(p$$-$$4)X+1$}.
\def\bun{\hbox{\bf b$_1$}}
This polynomial is also the characteristic polynomial of the
matrix associated to the rules~(\thematWB). 
The polynomial has two positive roots, the greatest one
is $\beta=\displaystyle{{\hbox{$(p$$-$$4)$}+\sqrt{\hbox{$(p$$-$$4)^2$}-4}}\over 2}$.
As $p\geq 7$, \hbox{$(p$$-$$4)^2>4$}, so that $\beta$ is a real number and 
$\beta>1$ as \hbox{$\displaystyle{{\hbox{$p$$-$4}\over2} > 1}$}. We easily check
that \hbox{$p$$-$$5 < \beta < p$$-$$4$} as \hbox{$P(p-5)=6$$-$$p \leq -1$} and as
\hbox{$P(p$$-$$4)=1$}. We set \hbox{$b= p$$-$4} and $\bun=b$$-$1.
It is known, see~\cite{fraenkel,frougny,hollander}, that we can represent any 
natural number~$n$ as a sum
of terms of the sequence defined by equation~(\therecseq)~:
\vskip 5pt
\ligne{\hfill
$\displaystyle{n=\sum\limits_{i=1}^k\alpha_i}$, where $\alpha_i\in\{0..\bun\}$.
\hfill\numlaform\hskip 20pt}
\newcounter{metalrep}
\setcounter{metalrep}{0}
\addtocounter{metalrep}{\value{laform}}
\vskip 5pt
This sum can be formally
represented in a position numeral system by \hbox{$\alpha_k..\alpha_1\alpha_0$}.
We shall write: \hbox{$[n]=\alpha_k..\alpha_1\alpha_0$}, where the $\alpha_i$'s 
are defined by~(\themetalrep) and we shall write
\hbox{$[\alpha_k..\alpha_1\alpha_0]=[[n]]=n$} for the converse operation.
Note that the representation is not unique. Indeed:

\begin{lemm}\label{notunique}
For all natural numbers $n$ and $k$,
\vskip 5pt
\ligne{\hfill
$(p$$-$$5)u_{n+k}+(p$$-$$5)u_n+
\displaystyle{\sum\limits_{i=1}^{k-1}\hbox{$(p$$-$$6)$}u_{n+i}}=u_{n+k+1}+u_{n-1}$
\hfill{\rm\numlaform}\hskip 20pt}
\newcounter{repshort}
\setcounter{repshort}{0}
\addtocounter{repshort}{\value{laform}}
\end{lemm}

\noindent
Proof of Lemma~\ref{notunique}. Using (\therecseq) we get:

$(p$$-$$5)u_{n+k}+(p$$-$$6)u_{n+k-1}= (p$$-$$4)u_{n+k}-u_{n+k}+(p$$-$$6)u_{n+k-1}$

\hskip 25pt $= u_{n+k+1} -u_{n+k} + (p$$-$$5)u_{n+k-1}$

So, that:

$(p$$-$$5)u_{n+k}+(p$$-$$6)u_{n+k-1}+ (p$$-$$6)u_{n+k-2} $

\hskip 25pt $= u_{n+k+1} -u_{n+k} + (p$$-$$5)u_{n+k-1} + (p$$-$$6)u_{n+k-2}$

\hskip 25pt $= u_{n+k+1} -u_{n+k} + u_{n+k} - u_{n+k-1} + (p$$-$$5)u_{n+k-2}$

\hskip 25pt $= u_{n+k+1} - u_{n+k-1} + (p$$-$$5)u_{n+k-2}$

By induction, on $i$ we get that :

\ligne{\hfill
$(p$$-$$5)u_{n+k} + \displaystyle{\sum\limits_{i=1}^{k-1}\hbox{$(p$$-$$6)$}u_{n+i}}
= u_{n+k+1} -u_{n+2} + (p$$-$$5)u_{n+1}$
\hfill($\ast$)\hskip 20pt}

Adding $(p$$-$$5)u_n$ to both sides, we get:

$(p$$-$$5)u_{n+k} + \displaystyle{\sum\limits_{i=1}^{k-1}\hbox{$(p$$-$$6)$}u_{n+i}}
+(p$$-$$5)u_n$

\hskip 25pt $= u_{n+k+1} -u_{n+2} + (p$$-$$5)u_{n+1} + (p$$-$$5)u_n$

\hskip 25pt $= u_{n+k-1} - ((p$$-$$4)u_{n+1}-u_n) + (p$$-$$5)u_{n+1} + (p$$-$$5)u_n$

\hskip 25pt $= u_{n+k+1} - u_{n+1} + (p$$-$$4)u_n= u_{n+k+1} + u_{n-1}$
\hfill\boxempty
\vskip 10pt
\def\bii{\hbox{\bf b$_2$}}
As $(p$$-$$5)u_n = (p$$-$$6)u_n+u_n$ we can rewrite~(\therepshort) as:

$(p$$-$$5)u_{n+k} +
\displaystyle{\left(\sum\limits_{i=0}^{k-1}\hbox{$(p$$-$$6)$}u_{n+i}\right)}+u_n
=u_{n+k+1}+u_{n-1}$

\noindent
making $n=0$ in the latter equation, as $u_{-1}=0$ and $u_0=1$ and replacing~$k$
by~$n$ this provides us with
\vskip 5pt
\ligne{\hfill
$(p$$-$$5)u_{n} +
\displaystyle{\left(\sum\limits_{i=0}^{n-1}\hbox{$(p$$-$$6)$}u_{i}\right)}
=u_{n+1}-1$\hfill\numlaform\hskip 20pt}
\newcounter{unmoinsun}
\setcounter{unmoinsun}{0}
\addtocounter{unmoinsun}{\value{laform}}

Again, using that $u_0=1$ we can again write:
\vskip 5pt
\ligne{\hfill
$(p$$-$$5)u_{n} +
\displaystyle{\left(\sum\limits_{i=1}^{n-1}\hbox{$(p$$-$$6)$}u_{i}\right)}
+ (p$$-$$5)u_0
=u_{n+1}$\hfill\numlaform\hskip 20pt}
\newcounter{unbis}
\setcounter{unbis}{0}
\addtocounter{unbis}{\value{laform}}

From~(\theunbis), replacing $n$ by~$n$+$k$ and taking a part under the sign~$\sum$,
we get:

\ligne{\hfill
$(p$$-$$5)u_{n+k} + \displaystyle{\sum\limits_{i=1}^{k-1}\hbox{$(p$$-$$6)$}u_{n+i}}
< u_{n+k+1}$
\hfill\numlaform\hskip 20pt}
\newcounter{forposit}
\setcounter{forposit}{0}
\addtocounter{forposit}{\value{laform}}


Note that (\theunbis) gives us another proof that the representation~(\themetalrep)
is not unique. Together with~(\theunmoinsun), we obtain that we necessarily go 
to $u_{n+1}$ just after the number given by the left-hand side of~(\theunmoinsun).
We have : $[u_{n+1}$$-$$1]=\bun\bii^n$. We can also say from~(\theunbis) that
the string \bun\bii$^k$\bun is ruled out for any $k\in\mathbb{N}$. 

Also note that the representation whose number of digits is greater is unique.
It is obtained by the following algorithm:
\vskip 5pt
\ligne{\hfill
\vtop{\leftskip 0pt\parindent 0pt\hsize=220pt
\ligne{{\sc input}: $n$ and $u_i$ for $i<n$, a table $a$;\hfill}
\ligne{\bf while $u_i \leq n$ loop $i := i$$+$1; end loop;\hfill}
\ligne{\bf for $j$ in reverse $\{0..i$$-$$1\}$\hfill}
\ligne{\bf loop $a_j := n$ div $u_j$; $n := n$ mod $u_j$; end loop;\hfill}
\ligne{{\sc output}: $a$ and $i$, the length of the table.}
}
\hfill}
\vskip 5pt
From these last 
remarks we get: 

\begin{thm}\label{ratcoord}{\rm(Margenstern, see~\cite{mmbook1})}
The language of the coordinates for the tessellations $\{p$$-$$2,4\}$ and $\{p,3\}$
with $p\geq 7$ is rational.
\end{thm}
We have that $[(p$$-$$5)u_{n+k} 
+ \displaystyle{\sum\limits_{i=1}^{k-1}\hbox{$(p$$-$$6$}u_{n+i}} +(p$$-$$5)u_n]
= \bun\bii^{k-1}\bun${\bf 0}$^n$, where $\bii=p$$-$6 and that 
$[u_{n+k+1}+u_{n-1}]= \hbox{\bf 10$^{k+1}$10$^{n-1}$}$.
The second representation has one more digits than the first one. 

From
now on, we take as coordinate of~$\nu$ the representation $[\nu]$ with the greatest
number of digits and we again denote it by~$[\nu]$.
\vskip 5pt
By induction, we define $U_n$ by 
\vskip 5pt
\ligne{\hfill
$U_0=u_0$ and $U_{n+1} = U_n+u_{n+1}$\hfill\numlaform\hskip 20pt
}
\newcounter{nbTreen}
\setcounter{nbTreen}{0}
\addtocounter{nbTreen}{\value{laform}}
\vskip 5pt
For our further study, we need a few results on the $u_n$'s and on the $U_n$'s.

\begin{lemm}\label{incrseq}
The sequences $\{u_n\}$ and $\{U_n\}$ are both increasing. For all positive~$n$
\vskip 5pt
\ligne{\hfill
$(p$$-$$5)u_n < u_{n+1} < (p$$-$$4)u_n$\hfill{\rm\numlaform}\hskip 20pt}
\newcounter{estnblev}
\setcounter{estnblev}{0}
\addtocounter{estnblev}{\value{laform}}
\vskip 5pt
\ligne{\hfill
$U_n+(p$$-$$6)u_n < u_{n+1} < U_n+(p$$-$$5)u_n$\hfill{\rm\numlaform}\hskip 20pt}
\newcounter{estnbtree}
\setcounter{estnbtree}{0}
\addtocounter{estnbtree}{\value{laform}}
\vskip 5pt
\ligne{\hfill
$u_{n+1} < U_{n+1} - u_n$\hfill{\rm\numlaform}\hskip 20pt}
\newcounter{estunplusun}
\setcounter{estunplusun}{0}
\addtocounter{estunplusun}{\value{laform}}
\end{lemm}

\noindent
Proof of Lemma~\ref{incrseq}.
From the equations (\therecWB) we derived from rules~(\thematWB), we can see that
the $v_n$'s and $w_n$'s are positive when $n>0$. This proves that the sequences
$\{v_n\}$ and $\{w_n\}$ are increasing which proves that the sequences $\{u_n\}$
and $\{U_n\}$ are also increasing. Summing the last two rows in (\therecWB) we
get that \hbox{$u_{n+1}= (p$$-$$4)w_n+(p$$-$$5)v_n$}, whence the inequalities
(\theestnblev).

We prove the left-hand side inequality of~(\theestnbtree) by induction:

$u_1 > U_0+ (p$$-$$6)u_0$ as $u_0=U_0=1$ and as $u_1=p$$-$4. Accordingly,
assume that $u_{n+1}> U_n+(p$$-$$6)u_n$. Then, using the second equation of~(\therecseq):
 
$u_{n+2}=(p$$-$$4)u_{n+1}-u_n = (p$$-$$6)u_{n+1}+u_{n+1}+u_{n+1}-u_n$

\hskip 25pt $> U_n + (p$$-$$6)u_n+(p$$-$$6)u_{n+1}+u_{n+1}-u_n$, by induction,

\hskip 25pt $= U_n+u_{n+1}+(p$$-$$6)u_{n+1}+(p$$-$$7)u_n$, so that, as $p\geq7$,

\hskip 25pt $\geq U_{n+1}+(p$$-$$6)u_{n+1}$.

The right-hand side of~(\theestnbtree) is easier: as $u_n < U_n$ when $n\geq 1$,

   $u_{n+1} < u_{n+1} + U_n -u_n < (p$$-$$4)u_n + U_n - u_n= U_n+(p$$-$$5)u_n$ 
from (\theestnblev), so that we get the proof of~(\theestnbtree).
From the definition of~$U_{n+1}$ and from $u_n < U_n$ when $n\geq 1$ we easily 
get~(\theestunplusun). 
\hfill\boxempty

   We need a similar lemma to Lemma~\ref{incrseq} when, using the same rules 
as~(\thematWB) we consider a 
tree~${\cal T}_B$
whose root is a $B$-node. Denote by~$y_n$ the number of nodes on the level~$n$
and by $Y_n$ the number of nodes in~${\cal T}_B$ down to the level~$n$, that one 
being included.

We have:

\begin{lemm}\label{incrseqB}
The sequences $\{y_n\}$ and $\{Y_n\}$ are both increasing. For all positive~$n$
\vskip 5pt
\ligne{\hfill
$Y_{n+1} = Y_n+y_{n+1}$\hfill{\rm\numlaform}\hskip 20pt}
\newcounter{nbtreeB}
\setcounter{nbtreeB}{0}
\addtocounter{nbtreeB}{\value{laform}}
\vskip 5pt
\ligne{\hfill
$u_{n+1} = (p$$-$$5)u_n+y_n$\hfill{\rm\numlaform}\hskip 20pt}
\newcounter{nblevB}
\setcounter{nblevB}{0}
\addtocounter{nblevB}{\value{laform}}
\vskip 5pt
\ligne{\hfill
$y_{n+1} = u_{n+1}-u_n$\hfill{\rm\numlaform}\hskip 20pt}
\newcounter{nbbfromB}
\setcounter{nbbfromB}{0}
\addtocounter{nbbfromB}{\value{laform}}
\vskip 5pt
\ligne{\hfill
$Y_{n+1} = U_{n+1}-U_n$\hfill{\rm\numlaform}\hskip 20pt}
\newcounter{nbtreefromB}
\setcounter{nbtreefromB}{0}
\addtocounter{nbtreefromB}{\value{laform}}
\end{lemm}

\noindent
Proof of Lemma~\ref{incrseqB}.
Formula~(\thenbtreeB) comes from the definitions of~$Y_n$ and of~$y_n$. As the 
rules~(\thematWB) also apply to the nodes of~${\cal T}_B$ formulas~(\therecWB)
also apply after replacing the first condition by $y_0=1$ and $y_1=4$.
Formula~(\thenblevB) comes from the decomposition of a tree down to the level~$n$+1 rooted 
at a $W$-node into $p$$-$5 copies of the same tree down to the level~$n$ and a copy
of ${\cal T}_B$ down to the level~$n$ thanks to the second rule of~(\thematWB).
Now, from~(\thenblevB) and from~(\therecseq) we get~(\thenbbfromB). At last,
from the definition of~$U_n$ and from~(\thenbtreeB) we get (\thenbtreefromB) by
induction. 
\hfill\boxempty

Now, we can transform (\theunbis) into:
\vskip 5pt
\ligne{\hfill
$u_{n+1} = U_n + 1 + (p$$-$$6)u_n 
+ \displaystyle{\sum\limits_{k=0}^{n-1}(\hbox{$p$$-$7})u_k}$
\hfill\numlaform\hskip 20pt}
\newcounter{unparL}
\setcounter{unparL}{0}
\addtocounter{unparL}{\value{laform}}
\vskip 5pt
\setbox210=\hbox{$(p$$-$$7)$}
\setbox211=\hbox{$(p$$-$$6)$}
Indeed, 

$U_n+1+(p$$-$$6)u_n+\displaystyle{\sum\limits_{k=0}^{n-1}\box210u_k}=
U_n+1+(p$$-$$6)u_n+\displaystyle{\sum\limits_{k=0}^{n-1}\copy211u_k}
-\displaystyle{\sum\limits_{k=0}^{n-1}}u_k$

$= U_n+1+(p$$-$$6)u_n+\displaystyle{\sum\limits_{k=0}^{n-1}\copy211u_k}-U_{n-1}=
u_n+1+(p$$-$$6)u_n+\displaystyle{\sum\limits_{k=0}^{n-1}\copy211u_k}$

$= (p$$-$$5)u_n+\displaystyle{\sum\limits_{k=1}^{n-1}\copy211u_k}+(p$$-$$5)u_0=u_{n+1}$

\noindent
according to~(\theunbis).
\vskip 5pt
On another side, the definition of~$U_n$ allows us to write:
\vskip 5pt
\ligne{\hfill
$u_{n+1} = U_{n+1} - \displaystyle{\sum\limits_{k=0}^nu_k}$1
\hfill\numlaform\hskip 20pt}
\newcounter{unparR}
\setcounter{unparR}{0}
\addtocounter{unparR}{\value{laform}}
\vskip 5pt
Formulas~(\theunparL) and~(\theunparR) will help us in the next Subsections.

\subsection{The preferred son approach}
\label{preferred}

   We now turn to a closer study of the tree, forgetting for a while the connection
with our tessellations. From this point, we shall denote a node either by its
number~$\nu$ or by its coordinate~$[\nu]$.

   The rules of~(\thematWB) indicates that each node has a single $B$-node.
The rules are a formal writing, they do not assign a place to the $B$-node
among the sons of the node. We may decide a particular display. For a reason which will be
later clear, let us consider this one:
\vskip 5pt
\ligne{\hfill
$\vcenter{
\vtop{\leftskip 0pt\parindent 0pt \hsize=85pt
\ligne{$\hbox to 15pt{$B$\hfill}\rightarrow\hskip 8pt W^{p-7}BW$\hfill}\vskip 0pt
\ligne{$\hbox to 15pt{$W$\hfill}\rightarrow\hskip 8pt W^{p-6}BW$\hfill}
}
}$
\hfill\numlaform\hskip 20pt}
\newcounter{prefWB}
\setcounter{prefWB}{0}
\addtocounter{prefWB}{\value{laform}}
\vskip 5pt

Denote by ${\cal P}_W$, ${\cal P}_B$ the tree which is obtained by a recursive
unlimited application of the rules~(\theprefWB) from a $W$-, $B$-node respectively. 
We call $W$-, $B$-\textbf{tree of height~$n$}, the sub-tree of~${\cal P}_W$,
${\cal P}_B$ respectively issued from the root,
down to the level~$n$, that level being included. 
Number the nodes of~${\cal P}_W$,
starting from the root to which we give number~1, and then go down level after level
and, on each level, from left to right. We identify a node with its number.
To each node~$\nu$ we associate the string~$[\nu]$ which we call the 
\textbf{coordinate} of the node. Define the \textbf{signature} of a node to be the 
lowest digit of its coordinate. Define, for a node, its \textbf{son signature} as the 
string which displays the signatures of its sons, from left to right.
We have the following property:

\begin{thm}\label{preftree}
In the tree~${\cal P}_W$, the son signature of any node is 
{\bf2$..$b$_\alpha$01}, with $p$$-$$4$, $p$$-$$5$ digits and $\alpha=1,\ 2$ for 
a $W$-, $B$-node respectively.
Consequently, the $B$-nodes are exactly those whose signature is~{\bf 0}. In each node,
a single son has a {\bf 0} signature: call it the {\bf preferred son} of the node.
In each node, the preferred son is the penultimate.
\end{thm}

Figure~\ref{tree9P} shows the first two levels of~$\cal P$. We can see that
the statement of Theorem~\ref{preftree} is observed in the figure. We also
call this tree the \textbf{preferred son tree}.
\vskip 5pt
\vtop{
\ligne{\hfill
\includegraphics[scale=0.6]{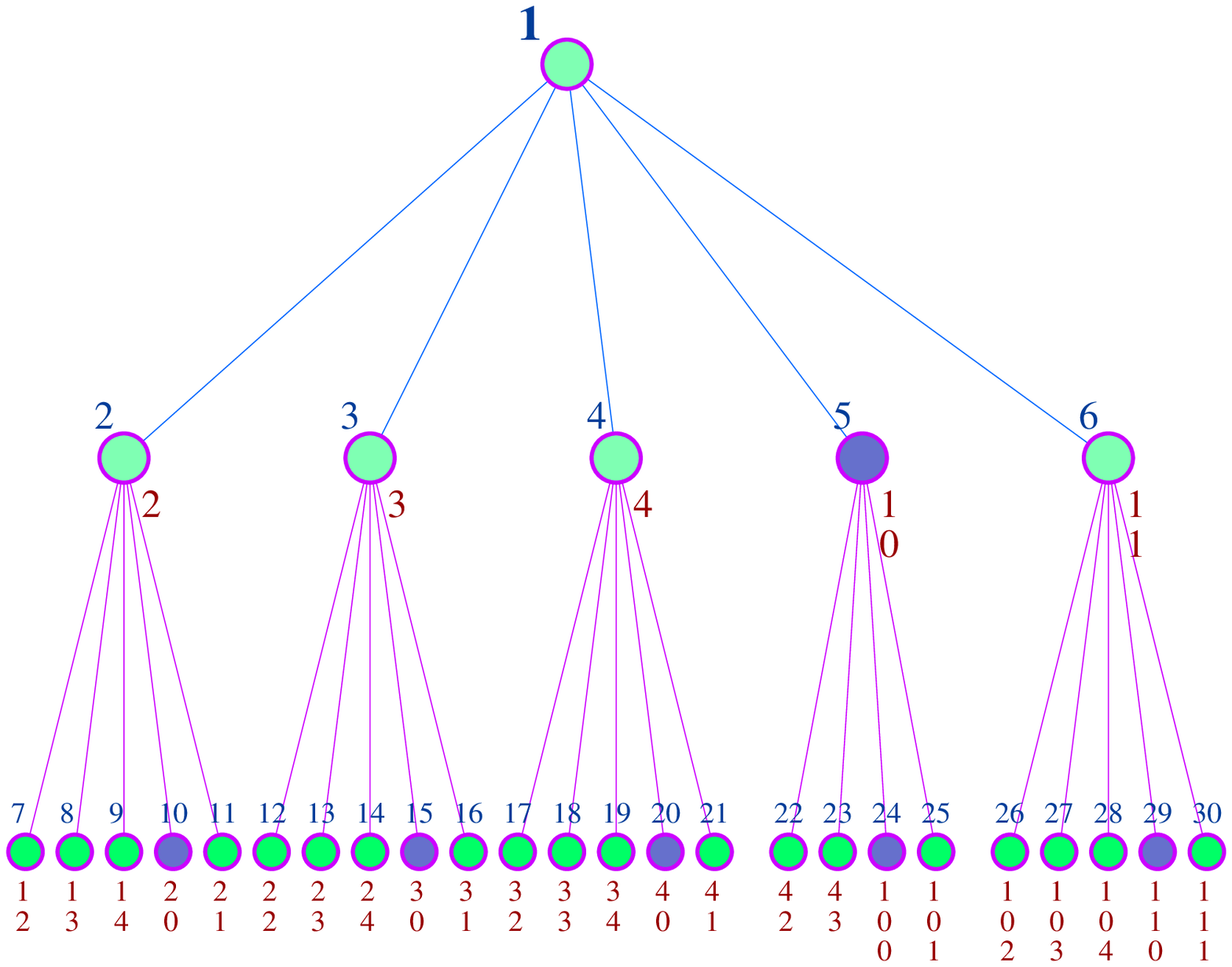}
\hfill}
\vspace{-25pt}
\begin{fig}\label{tree9P}
\leurre
The tree associated to the rules~{\rm(\theprefWB)}.
In that tree, the $B$-nodes are the preferred sons of the nodes.
For each node: in blue, the numbers, in red the coordinates.
\end{fig}
}
\vskip 5pt

\noindent
Proof of Theorem~\ref{preftree}. Figure~\ref{tree9P} shows us that the 
theorem is true for the root and for its sons, \textit{i.e.} for level~1.
As $U_n$ is the number of nodes up to the level~$n$, that level being included,
$U_n$ is also the number of the last node on the level, we also say the rightmost one,
and its coordinate is \textbf{1$^{n+1}$}. As a consequence, the coordinate of the first
node of the level~$n$ is \textbf{1$^{n-1}$2}. Indeed, from~(\theestnbtree) 
and~(\theestunplusun), we get that \hbox{$U_n<u_{n+1}<U_{n+1}$} so that
$[\nu]$ has $n$ digits when $\nu<u_n$ and $n$+1 digits when $\nu\geq u_n$.
Next, from~(\theestnbtree) and~(\theestunplusun) again, we obtain that $u_{n+1}$
always belong to the $B$-tree rooted at the $B$-son of the root.
Formulas~(\theunparL) and~(\theunparR) allow us to precisely locate~$u_{n+1}$ in
that tree. Indeed, consider formula~(\theunparR) with~$n=0$. It says $u_1$ is the
penultimate son of the root. If $n=1$, it says that~$u_2$ is the penultimate
node of the $B$-tree rooted at~$u_1$. By induction, the formula says that
$u_{n+1}$ is the penultimate node of the $B$-tree rooted at~$u_n$.
Formula~(\theunparL) says the same thing: instead of starting from~$U_{n+1}$
the last node on the level~$n$, we start from~$U_n$+1, the first node on that level.
The formula says that we cross the $p$$-$6 $W$-trees rooted at the first $p$$-$6 sons
of the root of~$\cal P$, and then we cross the $p$$-$7 $W$-trees of height $n$$-$1
rooted at the first $p$$-$7 $W$-sons of~$u_1$, then the $p$$-$7 $W$-trees of height 
$n$$-$2 rooted at the~$p$$-$7 $W$-sons of~$u_2$ and so on,
until we arrive at~$u_n$: we cross its $p$$-$7 $W$-sons before arriving to~$u_{n+1}$.
As the coordinate of $u_{n+1}$ is \textbf{10$^n$} by definition, the signature of
all $u_k$'s for $k\geq 1$ is \textbf{0} and, by the rules~(\theprefWB), they are
$B$-nodes.
Call the sequence of nodes $\{u_i\}_{i\in\mathbb{N}^+}$, where $\mathbb{N}^+$ denote
the set of positive integers, the \textbf{main $B$-line} of the tree ${\cal P}_W$.

\ifnum1=0{
Note that by our convention, the next node after node~$\nu$ when going to the last 
node of the level is $\nu$+1, except if $\nu$ is the last node on that level.
The position of the nodes with signature~\textbf{0} corresponds to a situation
when going from~$\nu$ to $\nu$+1 generates a carry. Remark that the son signature
of the root is \textbf{2$..\bun$01} and that it is also the case for the first node
on the level~2, of the $\cal P$, node~2. Now, except when $p=7$, node~3 is also a $W$-node 
and it has $p$$-$5 sons, so that its signature is also \textbf{2$..\bun$01}.
Say that the carry occurs when, going from~$\nu$ to~$\nu$+1, that latter node has
a \textbf{0}-signature. 
} \fi

Fix a node~$\nu$ in~${\cal P}_W$. Let ${\cal P}_\nu$ denote the sub-tree of~${\cal P}_W$
rooted at~$\nu$: we have that \hbox{${\cal P}_W={\cal P}_1$}.
Call \textbf{$B$-thread} of~$\nu$ the sequence of $B$-nodes 
$\{\nu_i\}_{i\in \mathbb{N}^+}$ such that $\nu_1$ is the $B$-son of~$\nu$  
and $\nu_{i+1}$ is the 
$B$-son of~$\nu_i$ for all $i\in\mathbb{N}$. Say that a node~$\mu$ is in the left-,
right-hand side of the $B$-thread if it is in a sub-tree of~${\cal P}_\nu$ rooted at 
one of the $W$-sons of~$\nu$ which come before~$\nu_1$, at the $W$-son of~$\nu$
which comes after~$\nu_1$ respectively. We shall say that $\mu$ is on the left-,
right-hand side of~$\nu$ it it is on the left-right-hand side of its $B$-thread 
respectively.

The theorem is an easy corollary of the next lemma.


\newdimen\nodelarge\nodelarge=39pt
\def\nodeligne #1 #2 #3 #4 #5 #6 #7 #8 {%
\ligne{\hfill\small\hbox to 15pt{\bf #1\hfill} 
\hbox to \nodelarge{#2\hfill} 
\hbox to \nodelarge{#3\hfill} 
\hbox to \nodelarge{#4\hfill} 
\hbox to \nodelarge{#5\hfill} 
\hbox to \nodelarge{#6\hfill} 
\hbox to \nodelarge{#7\hfill} 
\hbox to \nodelarge{#8\hfill} 
\hfill}
}
\begin{lemm}\label{inherit}
Let $[\nu]${\bf a} be the coordinate of a node~$\mu$ with
{\bf a}$~\in [${\bf 0}$..\bun$$]$. Then the coordinates of the sons of~$\mu$
are given by the following table:
\vskip 5pt
\ligne{\hfill
\vtop{\leftskip 0pt\parindent 0pt\hsize=360pt
\nodeligne {j} {$[\nu]${\bf j$_1$2}} {$[\nu]${\bf j$_1$3}} {$[\nu]${\bf j$_1$i}} 
{$[\nu]${\bf j$_1$}\bii} {$[\nu]${\bf j$_1$}\bun} {$[\nu]${\bf j0}} {$[\nu]${\bf j1}}
\nodeligne {0} {$[\nu_1]${\bf \bun2}} {$[\nu_1]${\bf \bun3}} 
{$[\nu_1]${\bf \bun i}} 
{$[\nu_1]${\bf \bun}\bii} {$[\nu]${\bf 0}} {$[\nu]${\bf 1}} {}
}
\hfill}
\vskip 5pt
\noindent
where $\nu_1 = \nu$$-$$1$, {\bf j}$~\in [1$..\bun$]$ and {\bf j}$_1 =~${\bf j}$-$$1$. 
Recall that \bii~= $p$$-$$6$. The first line gives the coordinates of the sons of~$\mu$
when its signature is {\bf j} with \hbox{{\bf j} $\not=$ {\bf 0}}. The second line
gives the coordinates when the signature of~$\mu$ is {\bf 0}.
\end{lemm}

Proof of Lemma~\ref{inherit}.
We proceed by induction on the level~$n$ of the node~$\mu$. We note that the lemma is
true for the root and that it is also true for its sons as illustrated by
Figure~\ref{tree9P}. Assume that the lemma is true up to the level~$n$.
Let $\mu$ be the first node of the level~$n$+1. From what we know,
its  is [{\bf 1$^n$2}] which we write $[\nu]${\bf 2} with 
$\nu = [[${\bf 1}$^n$]]. The coordinate of the first son of~$\mu$ is
[{\bf 1$^{n+1}$2}] which is of the form $[\nu]${\bf 12}. We can easily see
that we also have that the coordinate of the $i$-th son of~$\mu$
is $[\nu]${\bf 1i}$^1$ with {\bf i}$^1 =$~{\bf i}+1, for \hbox{$i \leq p$$-$6}.
Hence, for the $p$$-$6-th son, the coordinate is $[\nu]${\bf 1\bun}. For the
next node, as $b$ is not a digit, the coordinate is $[\nu]${\bf 20},
so that the coordinate of the last son is $[\nu]${\bf 21} and for the first son of $\mu$+1,
the coordinate is $[\nu]${\bf 22}. Note that by its position among the sons of~$\mu$,
its son $[\nu]${\bf 20} is a $B$-node. 

Now, we can repeat this argument for the sons of $\mu$+1, $\mu$+2, until $\mu$+$p$$-$6. 
Accordingly, the coordinate of the first son of~$\mu$+$p$$-$5, which is a $B$-node is 
$[\nu]${\bf \bun2}. When we arrive to the $p$$-$7-th son of~$\mu$+$p$$-$5,
its coordinate is $[\nu]${\bf \bun\bii}, so that for the next node, the $B$-son
of~$\mu$+$p$$-$5, the coordinate is $[\nu$+1]{\bf 00}. From this, we can see that,
arriving to the last son of the node $\mu$+$p$$-$4, its coordinate is
$[\nu$+1]{\bf 11}. This node is the last node on the tree rooted at the father of~$\mu$
and of height~2. By induction of the height of the tree whose first node is~$\mu$,
the induction being allowed by the position of the $u_i$'s given by~(\theestnbtree) 
and~(\theestunplusun), which coincide with a $B$-node, by the recursive application
of the rules~(\theprefWB), we have that the rules apply to the tree~${\cal P}_2$.
We have proved the theorem for a tree of height~$n$ which is not rooted at~1.
The similar argument still holds with ${\cal P}_i$ with $i<p$$-$5. Accordingly,
we arrive at the first node on the level~$n$ of ${\cal P}_5$ which is a $B$-tree.
Now, if we consider the sons of~5, namely nodes~22, 23, 24 and~25, see Figure~\ref{tree9P}.
The first two nodes are $W$-nodes for which we can repeat the previous argument as
${\cal P}_{22}$ and ${\cal P}_{23}$ are trees with the height $n$$-$1. The
crossing at the level~$n$+1 of~${\cal P}_W$ of ${\cal P}_{24}$ which is a $B$-tree
corresponds to a situation we have already met in the previous trees of
height~$n$ and which is controlled by~(\theestnbtree). We remain with ${\cal P}_{25}$ 
which is again a $W$-tree,
so that the theorem is also checked there, up to the considered level. So that
we proved the theorem for ${\cal P}_5$. Now with ${\cal P}_6$ we have again a
$W$-tree of height~$n$, so that the theorem is also checked there. Consequently,
we proved the theorem for the nodes on the level~$n$+1.\hfill\boxempty
\vskip 5pt
An interesting consequence of the lemma is the following property:

\begin{thm}\label{locate}
In each tree, the $B$-thread and the $B$-main line coincide. Define the {\bf a}-slice 
of a tree ${\cal P}_\nu$ as the nodes which are on the right-hand side of the
{\bf a}$_1$-th son of~$\nu$ and on the left-hand side of its {\bf a}-th son
when {\bf a}~$\in [${\bf 2}..\bii$]$, where {\bf a}$_1 =$~{\bf a}$-$1. 
The {\bf 0}-slice is on the right-hand side of the penultimate son and on the
left-hand side of the last one, the {\bf 1}-slice is on the right-hand side of the 
last node
and on the left-and side of the first one, the \bun-slice, only present if $\nu$
is a $W$-node, is on the left-hand side of the penultimate node and on the right-hand side
of the \bii-th node. Then, if $[\nu] =$~{\bf a}$_k$..{\bf a}$_1${\bf a}$_0$,
then $\nu$ is in the {\bf a}$_i$-th slice of $[[${\bf a}$_k$..{\bf a}$_i]]$ when
$i\in [1..k]$. We have that {\bf a}$_0$ gives the position in the tree
rooted at $[[${\bf a}$_k$..{\bf a}$_1]]$.
\end{thm}

This provides us with an algorithm to locate the nodes, but we have to do
more in order to get the branch, in the tree leading from~1 to the node.
For this, we need a few easy lemmas:

\begin{lemm}\label{incrcoord}
Let {\bf a}$_k$..{\bf a}$_1${\bf a}$_0$ be the coordinate of a node~$\nu$.
The coordinate of $\nu$$+$$1$ is given by Algorithm~{\rm\ref{a_incrcoord}}. 
The algorithm is linear in the size of~$[\nu]$.
\end{lemm}
\vskip 5pt
\ligne{\hfill
\vtop{\leftskip 0pt\parindent 0pt\hsize=260pt
\begin{algo}\label{a_incrcoord}\leurre
Algorithm for incrementing a number on its representation.
\end{algo}
\ligne{\bf $i := 0$;\hfill}
\ligne{\bf while $i\leq k$ and a$_i = \bii$\hfill}
\ligne{\bf loop $i := i$$+$$1$; end loop;\hfill}
\ligne{\bf if $i = 0$ and a$_0 = \bun$\hfill}
\ligne{\bf \hskip 15pt then a$_0 :=$~0; $k := k$$+$$1$; a$_1 := 1$; \hfill}
\ligne{\bf \hskip 15pt elsif $i = 0$; -{}- {\rm then} a$_0 < \bii$\hfill}
\ligne{\bf \hskip 40pt then a$_0 :=$~a$_0$$+$$1$; \hfill}
\ligne{\bf \hskip 15pt elsif $i > k$ then a$_0 :=$~a$_0$$+$$1$; \hfill}
\ligne{\bf \hskip 15pt elsif a$_i < \bun$ then a$_0 :=$~a$_0$$+$$1$; \hfill}
\ligne{\bf \hskip 15pt else for $j$ in $\{0..i\}$ loop a$_i :=$~0; end loop; \hfill}
\ligne{\bf \hskip 40pt if $i < k$ then a$_{i+1} :=$~a$_{i+1}$$+$$1$; \hfill}
\ligne{\bf \hskip 50pt else $k := k$$+$$1$; a$_k :=$~1;\hfill}
\ligne{\bf \hskip 40pt end if;\hfill}
\ligne{\bf end if;\hfill}
}
\hfill} 

Proof of lemma~\ref{incrcoord}.
If $i = 0$ and {\bf a}$_0\not=\bii$, the body of the loop is never executed, so that after
the {\bf while}, nothing is changed and we must perform {\bf a$_0 :=$~a}$_0$+1 unless
{\bf a}$_0=\bun$, in which case we go from \bun{} to {\bf 10}.
If we are not in this case, the body of the loop was executed at least once.
The {\bf while} achieves to perform the body of the loop, either because 
$i>k$ or because {\bf a$_i = \bii$}. If
$i>k$, all digits are \bii, so that {\bf a}$_0$ must become \bun. The other possibility 
is that $i\leq k$ and so {\bf a}$_i \not= \bii$. If {\bf a}$_i<\bii$, we perform
{\bf a$_0 :=$~a}$_0$+1. If not, {\bf a}$_i=\bun$ and so all {\bf a}$_j$'s from~0 
to~$i$ must be {\bf 0} and {\bf a}$_{i+1}$ becomes {\bf a}$_{i+1}$+1. Indeed,
as the pattern \bun\bun{} is ruled out, we had {\bf a}$_{i+1} < \bun$ before
performing this action. Nevertheless, this assumes
that $i<k$. If $i=k$, as {\bf a}$_k=\bun$, $k$ must be incremented by~1 and {\bf a}$_{k+1}$ 
must be~{\bf 1}.
\hfill\boxempty
\vskip 5pt
\noindent
\begin{lemm}\label{decrcoord}
Let {\bf a}$_k$..{\bf a}$_1${\bf a}$_0$ be the coordinate of a node~$\nu$,
assuming that $\nu\not=0$.
The coordinate of $\nu$$-$$1$ is given by Algorithm~{\rm\ref{a_decrcoord}}.
The algorithm is linear in the size of $[\nu]$.
\end{lemm}
\vskip-10pt
\ligne{\hfill
\vtop{\leftskip 0pt\parindent 0pt\hsize=260pt
\begin{algo}\label{a_decrcoord}\leurre
Algorithm for decrementing a number on its representation.
\end{algo}
\ligne{\bf $i := 0$;\hfill}
\ligne{\bf while a$_i =$~0\hfill}
\ligne{\bf loop a$_i := \bii$; $i := i$$+$$1$; end loop;\hfill}
\ligne{\bf if  $i =0$\hfill}
\ligne{\bf \hskip 15pt then a$_0 :=$~a$_0$$-$$1$;\hfill}
\ligne{\bf \hskip 15pt else a$_i :=$~a$_i$$-$$1$; a$_{i-1} := \bun$;\hfill}
\ligne{\bf end if;\hfill}
}
\hfill}
\vskip 5pt

\noindent
Proof of Lemma~\ref{decrcoord}.
Note that there is no need of a condition on $i$ with respect to~$k$ in the {\bf while}
as {\bf a$_k\not=$~0} according to the assumption that $\nu\not=0$. If $i = 0$
after the {\bf while}, the body of the loop was not executed, which means that 
{\bf a$_0\not=$~0}, so that {\bf a}$_0$ becomes {\bf a$_0$$-$$1$}. If $i$ is not~0
after the {\bf while}, we arrive at~{\bf a}$_i$ which is not~{\bf 0} and all
{\bf a}$_j$'s with $j<i$ are transformed into~\bii.  We can reduce {\bf a}$_i$ by~1
and, due to (\theunmoinsun), as $i>0$, we set {\bf a}$_{i-1}$ to \bun.
\hfill\boxempty

At last, the following lemma reminds us how to recognize the status of a node and
it allows also us to compute the coordinate of the father of a node. By convention,
the father of the root is~0. The lemma gives both the number and the coordinate.

\begin{lemm}\label{statfather}
Let $\nu$ be a node with {\bf $[\nu] =$~a$_k..$a$_1$a$_0$}. It is a $B$-node
if and only if {\bf a$_0 =$~0}. The father of $\nu$ is obtained as follows:
\vskip 5pt
If {\bf a$_0\in\{0,1\}$}, then the father is {\bf $[[$a$_k..$a$_1$$]]$}.
Otherwise, it is {\bf $[[[$a$_k..$a$_1$$]]+1]$}, where the latter number is computed
by the algorithm of Lemma~{\rm\ref{decrcoord}}.
\end{lemm}

\begin{lemm}\label{locdigit}
Consider an {\bf a}-slice in ${\cal P}_\nu$. Let $\nu_\ell$ and~$\nu_r$
be the sons of~$\nu$ which delimit the slice. Then the digits of the sons of~$\nu_\ell$
and of~$\nu_r$ which belong to the {\bf a}-slice are $[${\bf 1}..\bun$]$ if $\nu_r$
is a $W$-node. They belong to $[${\bf 1}..\bii$]$ if $\nu_r$ is a $B$-node.
\end{lemm}

The lemma allows us to prove the correctness of Algorithm~\ref{a_linalgo} stated in 
the next theorem.
The algorithm constructs a path going from the root to the node, constituted by 
new digits in $\{$1..$p$$-$$4\}$. The first digit~$d_1$ indicates the path from the 
root to the $d_1$-th son of the root. If we arrived at a node~$\nu$ through
$d_1..d_i$, $d_{i+1}$ indicates the $d_{i+1}$-th node of~$\nu$. Clearly, 
$p$$-$4 is never used for a $B$-node and, the digit~$p$$-$5 leads to the $B$-son
in a $W$-node while in a $B$-node, the $B$-son is reached through the digit~$p$$-$6.

\vskip 5pt
\ligne{\hfill
\vtop{\leftskip 0pt\parindent 0pt\hsize=260pt
\begin{algo}\label{a_linalgo}\leurre
Algorithm to compute the path from the root to the node when given its coordinate.
\end{algo}
\ligne{{\sc input}: {\bf a$_k..$ a$_0$}; $\ell$, $r$: tables of size $k$$+$$1$;\hfill} 
\ligne{\hskip 35pt $prev := k$; $s_\ell := W$, $s_r := W$;\hfill}
\ligne{\bf \hskip 35pt procedure actualize ($a$, $b$, $i$, $t$) is\hfill}
\ligne{\bf \hskip 35pt begin\hfill}
\ligne{\bf \hskip 50pt for j in $[i..t]$\hfill}
\ligne{\bf \hskip 50pt loop $a(j) := b(j)$; end loop; $t := i$;\hfill}
\ligne{\bf \hskip 35pt end procedure;\hfill}
\ligne{{\bf for $i$ in reverse $[0..k]$}\hfill}
\ligne{\bf loop if a$_i$ in $[$2$..$\bun$]$\hfill}
\ligne{\bf \hskip 35pt then $\ell(i) :=$~a$_i$$-$$1$; $r(i) :=$~a$_i$;\hfill}
\ligne{\bf \hskip 60pt if $((s_\ell = B)$ and $($a$_i=\bii$$))$\hfill}
\ligne{\bf \hskip 75pt or $((s_\ell = W)$ and $($a$_i=\bun$$))$\hfill}
\ligne{\bf \hskip 70pt then $s_\ell := W$; $s_r := B$;\hfill}
\ligne{\bf \hskip 70pt else $s_\ell := W$; $s_r := W$;\hfill}
\ligne{\bf \hskip 60pt end if;\hfill}
\ligne{\bf \hskip 60pt actualize($\ell,r,i$$+$$1,prev$);\hfill}
\ligne{\bf \hskip 26pt elsif a$_i =$~0\hfill}
\ligne{\bf \hskip 35pt then if $s_\ell := B$;\hfill}
\ligne{\bf \hskip 70pt then $\ell(i) := p$$-$$6$; $r(i) := p$$-$$5$;\hfill}
\ligne{\bf \hskip 70pt else $\ell(i) := p$$-$$5$; $r(i) := p$$-$$4$;\hfill}
\ligne{\bf \hskip 92.5pt $s_\ell := B$; $s_r := W$;\hfill}
\ligne{\bf \hskip 60pt end if;\hfill}
\ligne{\bf \hskip 60pt actualize($r,\ell,i$$+$$1,prev$);\hfill}
\ligne{\bf \hskip 26pt elsif a$_i =$~1\hfill}
\ligne{\bf \hskip 35pt then if $i=k$\hfill}
\ligne{\bf \hskip 70pt then $\ell(i) := 0$; $r(i) := 0$;\hfill}
\ligne{\bf \hskip 70pt else if $s_\ell = B$\hfill}
\ligne{\bf \hskip 105pt then $\ell(i) := p$$-$$5$;\hfill}
\ligne{\bf \hskip 105pt else $\ell(i) := p$$-$$4$;\hfill}
\ligne{\bf \hskip 95pt end if;\hfill}
\ligne{\bf \hskip 95pt $r(i) := 1$; $s_\ell := W$; $s_r := W$;\hfill}
\ligne{\bf \hskip 60pt end if;\hfill}
\ligne{\bf \hskip 26pt end if;\hfill}
\ligne{\bf end loop;\hfill}
\ligne{{\sc output}: $\ell$; \hfill}
}
\hfill}
\vskip 5pt

\begin{thm}\label{linalgo}{\rm(Margenstern, see~\cite{mmbook1})}
Algorithm~{\ref{a_linalgo}} computes the branch from the root of~${\cal P}_W$ to
a node~$\nu$ in $[\nu]$. The algorithm is linear in the length of $[\nu]$.
\end{thm}

\noindent
Proof of Theorem~\ref{linalgo}.
The idea of the algorithm is to use the property stated by Lemma~\ref{locdigit}.
Each digit of the coordinate indicates in which slice of the tree the node occurs. 
Note that a slice is delimited by two nodes, say the 
\textbf{left-, right-hand side milestones}. The problem is to identify which
milestone belongs to the path, in the tree, leading from the root to the node.

We construct the path by using two tables, $\ell$ and~$r$: $\ell$, $r$ contains the path 
from the root to the left-, right-hand side milestone of the current slice respectively.
As indicated before, the path consists in directions encoded by digits in 
$\{1..p$$-$$4\}$. Assume that the paths
are $\ell$ and~$r$ for the left-, right-hand side milestone of the current slice 
respectively. Let {\bf a} be the new digit.

If {\bf a~$\in\{$2$..\bun\}$}, then the left-hand side milestone is a son of the
right-hand side one at the previous step. So it is also the case for the new 
right-hand side milestone. Eventually, we have to copy $r$ onto~$\ell$ from the last
index until which the contents of tables were equal. If {\bf a $=$ 0}, then,
the left-hand side milestone is the $B$-son of the previous left-hand side 
milestone~$\ell_0$, and the right-hand side milestone is the last son of~$\ell_0$.
We have to eventually copy a part of~$\ell$ onto~$r$. If {\bf a $=$ 1}, then
the new left-hand side milestone is the last son of the previous one while the
new right-hand side milestone is the first son of the previous one. In that case, there is
no actualization but there will be later. Note that the node whose coordinate is
defined by the digits already visited is given by the path defined by the content
of~$\ell$ down to this digit.
This leads us to Algorithm~\ref{a_linalgo}.

\noindent
This completes the proof of Theorem~\ref{linalgo}. \hfill\boxempty

\ifnum 1=0 { 
\begin{lemm}\label{prefcoord}
Let $\nu$ be a node of~${\cal P}$, different from the root of that tree.
Let $\{\nu_i\}_{i\in \mathbb{N}}$ be the $B$-line of~$\nu$. We have that 
$[\nu_i]=[\nu]${\bf 0}$^i$ for all $i\in\mathbb{N}$. Let $\mu$ be a node 
of~${\cal P}_\nu$, $\mu\not=\nu$. If $\mu$ is in the left-hand side of the $B$-line
of~$\nu$, $[\mu]=[\nu$$-$$1]\alpha$, if $\mu$ is in the right-hand side of the
$B$-line of~$\nu$, $[\mu]=[\nu]\beta$, where $\alpha$ and $\beta$ are strings
on \hbox{\bf 0$..\bun$}.
\end{lemm}

\noindent
Proof of Lemma~\ref{prefcoord}. We prove the theorem for all trees at the same time,
by induction on the levels.
We note that the theorem holds for the levels~0 and~1 of $\cal P$.
Assume that it is true for all nodes down to the level~$n$.

Consider the level~$n$+1. Let $\nu$ be its first node. We know that $\nu=U_n+1$.
The first node's coordinate is [\textbf{1$^n$2}]. 
The coordinate of the father~$\mu$ of that node is [\textbf{1$^{n-1}$}2] so that the
statement of the theorem is true. The node~$\mu$ has $p$$-$4 sons.
The signature of the first one is~{\bf 2}, so that the signature of the $(p$$-$$6)$-th son 
is $\bun$. Accordingly, the next one should be $b=u_1$. Now, this $u_1$ is
appended to the one which corresponds to $\mu$$-1$'s coordinate. Accordingly,
the coordinate of the $(p$$-$$5)$-th son of~$\mu$ is [\textbf{1$^{n-1}$20}] 
\textit{i.e.} it is $\mu${\bf 0}. But this node is also a $B$-node by its position among 
the sons of~$\mu$. The coordinate of the last son of~$\mu$ is then [\textbf{1$^{n-1}$21}]
which satisfies the theorem.

   This coordinate of the last son of $\nu$, shows that the hypothesis of the theorem
is satisfied for the first son of~$\nu$+1. Its coordinate is [\textbf{1$^{n-1}$22}].
The just mentioned argument can be repeated, so that we get that the signature
of the $B$-son of~$\nu$+1 is {\bf 0}, that its coordinate is [$\mu$+1]{\bf 0}.
We can repeat this argument until we meet the first $B$-son on the level~$n$
whose number is $\mu$+$p$$-$6. Hence, its coordinate is 
$[\mu$+$p$$-$$6]=[\hbox{\textbf{1$^{n-1}\bun$}}]$ and the coordinate of the first son of 
that node is [{\bf 1$^{n-1}\bun$2}]. The coordinate of the $(p$$-$$6)$-th son 
of~$[\mu$+$p$$-$$6]$ should be [\hbox{\textbf{1$^{n-1}\bun\bun$}}] but this writing
is not accepted: [$\bun\bun$] should be replaced by [{\bf 100}] that is~$u_2$
according to (\therepshort) from Lemma~\ref{notunique} when $n=k=0$. This
shows us that the coordinate of the $(p$$-$$6)$-th son of $[\mu$+$p$$-$$6]$
is [\hbox{\textbf{1$^{n-2}$200}}]. Now, As $\mu$+$p$$-$6 is a $B$-son, it
has $p$$-$5 sons, so that the coordinate of its last son is
[\hbox{\textbf{1$^{n-2}$201}}] whose signature is {\bf 1}.
Accordingly, when we cross all the sons of~$\mu$, we still
met the same son signatures at each node, taking into account the difference
between $B$- and $W$-nodes. Accordingly, we know that for the node~$\omega$
which comes after $\mu$+$p$$-$$5$, the signature of the first son is again~{\bf 2}.

From this
\noindent
} \fi

\subsection{The leftmost approach}
\label{leftmost}

   The leftmost approach is based on the decomposition we have seen in the
illustrations given by Figures~\ref{bijectionp4} and~\ref{bijectionp3} and
in the process described in the proofs of Theorems~\ref{treebijp4} and~\ref{treebijp3}.
It attaches the $B$-node to the strip so that if the orientation from left to right
is identified with a counter-clockwise motion around a point in the hyperbolic plane,
then the $B$-node is the leftmost son of a node, whence the title of the current section.  
We call this new tree the \textbf{leftmost son tree}.

The numbering of the nodes is the same as the rules define the same number of
$B$-sons and $W$-sons as in the Subsection~\ref{preferred}. We also keep the
same definition of the coordinates. The new tree is not exactly the same as the previous 
one: the branches are different, except in some part of the tree as we shall see.

Consider the representation of the tree given in Figure~\ref{tree9L}. We can check
on the figure that the root is a $W$-node and that each node has one $B$-son exactly.
Number the nodes starting from~1 which we assign to the root and then, level after level
and on each level, from left to right. 

In the figure, we also can remark the following property. The signature of some nodes
is~0, and it seems that each node has one son exactly whose signature is~0. This
property is true for each node of the tree as below stated. As the branches of the
tree are different from that of Subsection~\ref{preferred}, this requires a proof.

\begin{thm}\label{thpref}
Consider the leftmost son tree~$\cal T$, which is associated to a sector of the 
tessellations $\{p,3\}$ and \hbox{$\{p$$-$$2,4\}$} 
with $p\geq7$. The tree has two kinds of nodes, $B$- and $W$-nodes which
are distributed in the tree according to the rules~{\rm\thematWB}. 
Among the
sons of the node~$n$, exactly one has $[n]0$ as its coordinate. Call it again the
{\bf preferred} son of the node~$n$. In the $B$-nodes, the preferred son is the
rightmost one. Consider a $W$-node~$n$. Say that its {\bf type} is~$1$ if
the preferred son is the rightmost one, otherwise, say that it is~$2$. In any node, the
preferred son is always a $W$-node of type~$2$. The root of the tree is a $W$-node of 
type~$2$. In a $W$-node of type~$1$, all $W$-sons but the preferred one are of type~$1$.
In a $W$-node of type~$2$, all $W$-sons but two of them are of type~$1$. The two $W$-sons
of type~$2$ in a node of type~$2$ are its preferred son and its rightmost son.
\end{thm}

\vtop{
\ligne{\hfill
\includegraphics[scale=0.6]{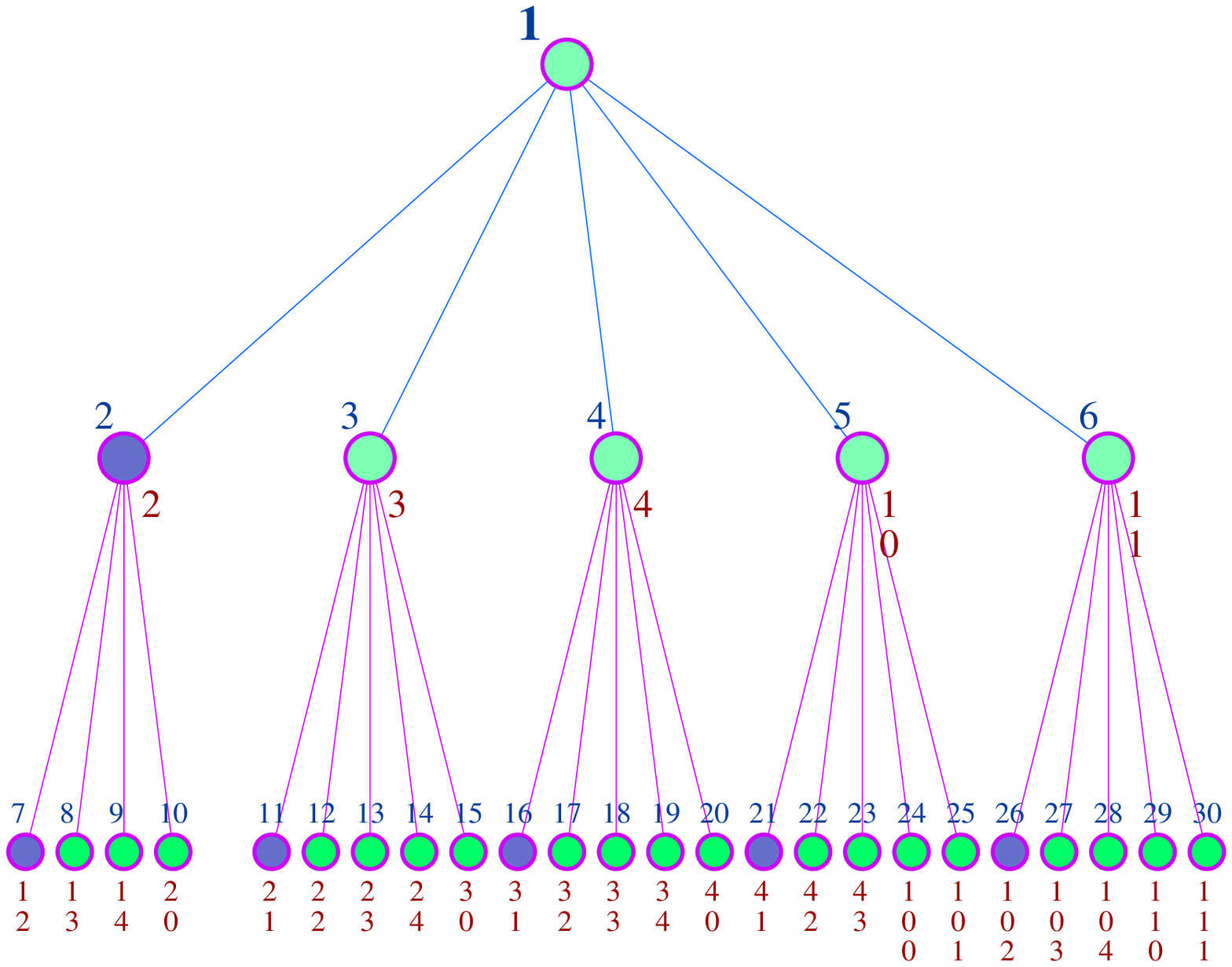}
\hfill}
\vspace{-25pt}
\begin{fig}\label{tree9L}
\leurre
The associated with the decomposition of a sector of the tessellations
as indicated in Figures~{\rm\ref{treebijp4}} and~{\rm\ref{treebijp3}}. In that
tree, the $B$-nodes are the leftmost son of a node. Here too, for each node, the number 
appears in blue, above the node, the coordinate is written in red, below, vertically.
\end{fig}
}
\vskip 5pt
\noindent
Proof of Theorem~\ref{thpref}.
As the nodes with signature~{\bf 0} are the same in both trees, we prove
the theorem by comparing the distributions of the sub-trees in each tree.
As the number of nodes are the same, we remark that the differences of distribution
appear inside the $W$-trees.

Let us compare the first level of the trees in Figures~\ref{tree9P} and~\ref{tree9L}.
It appears that the difference is a permutation operating on the first and the
$p$$-$5-th sons of the root which is a $W$-node in both cases. As a $B$-node has
one node less than a $W$-node, the above permutation entails a shift on the position of the
trees issued from the nodes of the first level compared to the similar trees in
the case of the preferred son display. Accordingly, up to~$u_2$, the preferred son
of the nodes~2 up to~$p$$-$5 being included is the rightmost son. Now, as
the position of the leftmost and rightmost branches of the tree issued from the last
node, which is a $W$-node in both trees, in that tree the preferred son is the penultimate
son, as it is in the preferred son tree. We can see that there are two kinds of $W$-nodes:
those we call of \textbf{type 1}, where the preferred son is the last one, and those
of \textbf{type 2} where the preferred son is the penultimate one. We note that the
the root is a $W$-node of type~2. Its last son is also of type~$2$. Now, its preferred
son~$\omega$ is now a $W$-node. Let $\cal W$ be 
$W$-tree issued from~$\omega$ in the leftmost son tree and let $\cal B$ be the 
$B$-tree issued from~$\omega$ in the preferred son tree.
From what we noticed on the trees issued from the last node of the root, $\cal W$
and $\cal B$ have the same rightmost branch. On the first level of $\cal W$ compared to
that  of $\cal B$, the additional node is outside $\cal B$. In particular, as the
preferred sons are the same in the preferred son and the leftmost son trees, the
preferred son of~$\omega$ has the same position with respect to the rightmost branch
of both~$\cal B$ and~$\cal W$. Accordingly, $\cal W$ is also of type~2. 
We also note that the leftmost and the preferred displays inside a $B$-tree shows us the 
same permutation between the $B$-sons, namely between the first one and the penultimate
one. Consequently, what we notice for $B$- and $W$-sons also holds for leftmost son
$B$-trees. The theorem is proved. \hfill\boxempty

In this context, we have a property which is analogous to that which is stated
in Theorem~\ref{linalgo}.
 
\begin{thm}\label{linalgoleft}{\rm(Margenstern, see~\cite{mmbook1})}
Algorithm~{\rm\ref{a_linalgoleft}} computes the branch from the root of~${\cal P}_W$ to
a node~$\nu$ of a leftmost son tree in $[\nu]$. The algorithm is linear in the length 
of $[\nu]$.
\end{thm}

\begin{lemm}\label{lastdigits}
The son signatures of the nodes in a leftmost son tree are the following ones:
\vskip 5pt
\ligne{\hfill$\vcenter{
\vtop{\leftskip 0pt\parindent 0pt\hsize=140pt
\ligne{\hbox to 80pt{$B$-node\hfill}\hskip 10pt {\bf 2$..b_1$0}\hfill}
\ligne{\hbox to 80pt{$W$-node of type~$1$\hfill}\hskip 10pt {\bf 1$..b_1$0}\hfill}
\ligne{\hbox to 80pt{$W$-node of type~$2$\hfill}\hskip 10pt {\bf 2$..b_1$01}\hfill}
\ligne{\hbox to 80pt{\hfill}\hskip 10pt {\bf 1$..b_2$01}\hfill}
}}$
\hfill{\rm\numlaform}\hskip 20pt}
\end{lemm}

\noindent
Proof of Lemma~\ref{lastdigits}.
The proof comes 
from the fact that the son signature is always {\bf 2$..b_1$01} in
preferred son trees and on the shift we observed in the proof of Theorem~\ref{thpref}.
By induction, assume that the lemma is true for the node~$\nu$. If $\nu$ is a $B$-node,
its signature is \hbox{\bf 2$..b_1$0}. Now, the node~$\nu$+1 is necessarily a $W$-node of 
type~1 and in its signature, the first digit is {\bf 1}. 
As a $W$-node has $p$$-$4 nodes,
the signature is \hbox{\bf 1$..b_1$0}. If $\nu$ is a $W$-node of type~1, its signature
is \hbox{\bf 1$..b_1$0}. Hence, the first digit of the signature of $\nu$+1 is {\bf 1}.
As $\nu$+1 is either
a $W$-node of type~1 or a $W$-node of type~2, its signature is \hbox{\bf 1$..b_1$0} or
\hbox{\bf 1$..b_2$01} respectively: in a $W$-node of type~2, the preferred son is
the penultimate.
If $\nu$ is a $W$-node of type~2, the last digit of its signature is \hbox{\bf 1}
as the preferred son is the penultimate. Hence, whatever $\nu$+1, either a $W$-node of 
type~2 or a $B$-node, the first digit of its signature is \hbox{\bf 2}.
Hence If $\nu$+1 is a $W$-node of type~2, a $B$-node, its signature is 
\hbox{\bf 2$..b_2$01}, \hbox{\bf 2$..b_2$0} respectively.
Accordingly, the signature of $\nu$+1 is one of the signatures indicated in the lemma
according to the status of the node.
\hfill\boxempty

\vskip-5pt
\ligne{\hfill
\vtop{\leftskip 0pt\parindent 0pt\hsize=340pt
\begin{algo}\label{a_linalgoleft}\leurre
Algorithm for computing the path from the root to a node in the leftmost son tree.
\end{algo}
\ligne{{\sc input}: {\bf a$_k..$ a$_0$}; $\ell$, $r$: tables of size $k$$+$$1$;\hfill} 
\ligne{\hskip 35pt $prev := k$; \hfill} 
\ligne{\hskip 35pt $s_\ell := 0$; $s_r := 0$; \hfill}
\ligne{\bf \hskip 35pt procedure actualize ($a$, $b$, $i$, $t$) is\hfill}
\ligne{\bf \hskip 35pt begin\hfill}
\ligne{\bf \hskip 50pt for j in $[i..t]$\hfill}
\ligne{\bf \hskip 50pt loop $a(j) := b(j)$; end loop; $t := i$;\hfill}
\ligne{\bf \hskip 35pt end procedure;\hfill}
\ligne{{\bf for $i$ in reverse $[0..k]$}\hfill}
\ligne{\bf loop if $s_\ell = 0$ and $s_r = 0$ and $i=k$\hfill} 
\ligne{\bf \hskip 35pt then if a$_i =$~1\hfill}
\ligne{\bf \hskip 70pt then $s_\ell := W_2$; $s_r := B$; 
$\ell(i):=0$; 
$r(i) :=0$; 
\hfill}
\ligne{\bf \hskip 70pt else $s_\ell := W_1$; $s_r := W_1$;
$\ell(i) :=$ a$_i$$-$$1$; $r(i) :=$ a$_i$; \hfill}
\ligne{\bf \hskip 95pt if a$_i = \bun$ then $s_r := W_2$; end if;\hfill}
\ligne{\bf \hskip 65pt end if;\hfill}
\ligne{\bf \hskip 35pt elsif $(s_\ell := B$ and $s_r := W_1)$
or $(s_\ell := W_1$ and $s_r = W_1)$\hfill}
\ligne{\bf \hskip 40pt or $(s_\ell = W_1$ and $s_r = W_2)$\hfill}
\ligne{\bf \hskip 35pt then if a$_i$ in $[$1$..$\bun$]$\hfill}
\ligne{\bf \hskip 70pt then $\ell(i) :=$~a$_i$; $r(i) :=$~a$_i$$+$$1$; 
$s_\ell := W_1$; \hfill}
\ligne{\bf \hskip 90pt if a$_i = \bii$ and $s_r = W_2$ then $s_r := W_2$; \hfill}
\ligne{\bf \hskip 105pt elsif a$_i < \bun$ then $s_r := W_1$; 
else $s_r := W_2$;\hfill} 
\ligne{\bf \hskip 95pt end if;\hfill}
\ligne{\bf \hskip 95pt actualize($\ell,r,i$$+$$1,prev$);\hfill}
\ligne{\bf \hskip 75pt else $r(i) := 1$; $\ell(i) := p$$-$$5$;\hfill}
\ligne{\bf \hskip 100pt if $s_\ell = W_1$ then $\ell(i) := \ell(i)$$+$$1$; end if;\hfill}
\ligne{\bf \hskip 92.5pt     $s_\ell := W_2$; $s_r := B$;\hfill};
\ligne{\bf \hskip 62pt end if;\hfill}
\ligne{\bf \hskip 35pt elsif $(s_\ell := W_2$ and $s_r := W_2)$
or $(s_\ell := W_2$ and $s_r = B)$\hfill}
\ligne{\bf \hskip 35pt then if a$_i$ in $[$2$..$\bun$]$\hfill}
\ligne{\bf \hskip 70pt then $\ell(i) :=$~a$_i$$-$$1$; $r(i) :=$~a$_i$; 
$s_\ell := W_1$; \hfill}
\ligne{\bf \hskip 95pt if a$_i < \bun$ then $s_r := W_1$; 
else $s_r := W_2$; 
end if;\hfill}
\ligne{\bf \hskip 95pt actualize($\ell,r,i$$+$$1,prev$);\hfill}
\ligne{\bf \hskip 75pt elsif a$_i =$~0 then $\ell(i) := p$$-$$5$; $r(i) := p$$-$$4$;\hfill}
\ligne{\bf \hskip 95pt $s_\ell := W_2$; $s_r := W_2$; \hfill}
\ligne{\bf \hskip 95pt actualize($r,\ell,i$$+$$1,prev$);\hfill}
\ligne{\bf \hskip 75pt else $\ell(i) := p$$-$$4$; $r(i) := 1$; 
$s_\ell := W_2$; $s_r := B$;\hfill};
\ligne{\bf \hskip 62pt end if;\hfill}
\ligne{\bf \hskip 26pt end if;\hfill}
\ligne{\bf end loop;\hfill}
\ligne{{\sc output}: $\ell$; \hfill}
}
\hfill}
\vskip 5pt
To prove the theorem, we need the following property:

It is interesting to pay a new visit to Theorem~\ref{linalgo} in the context of the
leftmost son tree. The theorem is still true in this new frame. We again state it
in this new context as the tree being different, the algorithm is also different.

\noindent
Proof of Theorem~\ref{linalgoleft}.
The basic point of the proof, namely Theorem~\ref{locate}, is true. Indeed,
the proof of that theorem lies on the notion of slice. As here $B$-nodes and node
with signature~{\bf 0} always being different, we replace the $B$-nodes 
by the preferred son: the \textbf{posterity} of a node~$\nu$ is a sequence
of nodes $\{\nu_i\}_{i\in\mathbb{N}}$ such that $\nu_0=\nu$ and $\nu_{i+1}$ is the
preferred son of~$\nu_i$. In the tree rooted at~$\nu$, the slices are again
the set of nodes between the posterities of two consecutive sons of~$\nu$. Now,
from this definition, we cans see that the slices are the same in a preferred son tree
and in a leftmost son one. The differences between the trees are the branches
in the slices. Now, what is changed with respect to the situation of 
Theorem~\ref{linalgo} is the delimitations of the sub-trees. We have six possible
situations, depending on the values of $s_\ell$ and~$s_r$, the type of node
of the current node. They are indicated by the following pairs:
\vskip 5pt
\ligne{\hfill
\hbox to 45pt{$0$ - $0$\hfill}
\hbox to 45pt{$B$ - $W_1$\hfill}
\hbox to 45pt{$W_1$ - $W_1$\hfill}
\hbox to 45pt{$W_1$ - $W_2$\hfill}
\hbox to 45pt{$W_2$ - $W_2$\hfill}
\hbox to 45pt{$W_2$ - $B$\hfill}
\hfill}
\vskip 5pt
Note that in the situation \hbox{$W_1$ - $W_2$}, the son signature of the $W_2$-node
is \hbox{\bf 12$..\bii$01}, that in the situation \hbox{$W_2$ - $W_2$},
the son signature of the first $W_2$-node is \hbox{\bf 12$..\bii$01} and that of the
second one is \hbox{\bf 2$..\bun$01}. At last, in the situation \hbox{$W_2$ - $B$},
the son signature of the $W_2$-node is \hbox{\bf 2$..\bun$01}. 
Note that \hbox{$0$ - $0$} is the situation of the beginning of the process.

As in the case of the preferred son tree, we construct two paths $\ell$ and~$r$. 
The current node, corresponding to the already examined digits, is always reached
by $\ell$.
This leads us to Algorithm~\ref{a_linalgoleft}.\hfill\boxempty

\section{The coordinates of a node and of its neighbours}
\label{res}

The preferred son and the leftmost trees are different but, as they both are sub-graphs
of the dual graph of the tiling, each node has the same neighbours, whichever the tree.

It is interesting to restore the dual graph from the tree. As the leftmost son tree
is tightly connected with the decomposition we introduced in Section~\ref{treebij}
in both tilings, it is not difficult to see that, in the
case of the tessellations~$\{p$$-$$2,4\}$, $\{p,3\}$, the $p$$-$2, $p$$-$3-th respectively
neighbour of a tile~$\nu$ is the first son of the tile~$\nu$+1. This can easily be
performed in the leftmost son tree: to each node, we append a connection between the
tile~$\nu$ and the first son of~$\nu$+1 which is the next node after the last son
of~$\nu$ on the level of its sons. For the case of the tessellations~$\{p,3\}$ there
are two other connections for~$\nu$: the connections with its neighbours on the same level,
namely the nodes~$\nu$$-$1 and~$\nu$+1. As the nodes are the same in both trees as well
as their dual graph, we can see that the connections are a bit more complex to be
established in the case of the preferred son tree. 

\vskip 10pt
\vtop{
\ligne{\hfill
\includegraphics[scale=0.33]{tree_9left.ps}
\includegraphics[scale=0.33]{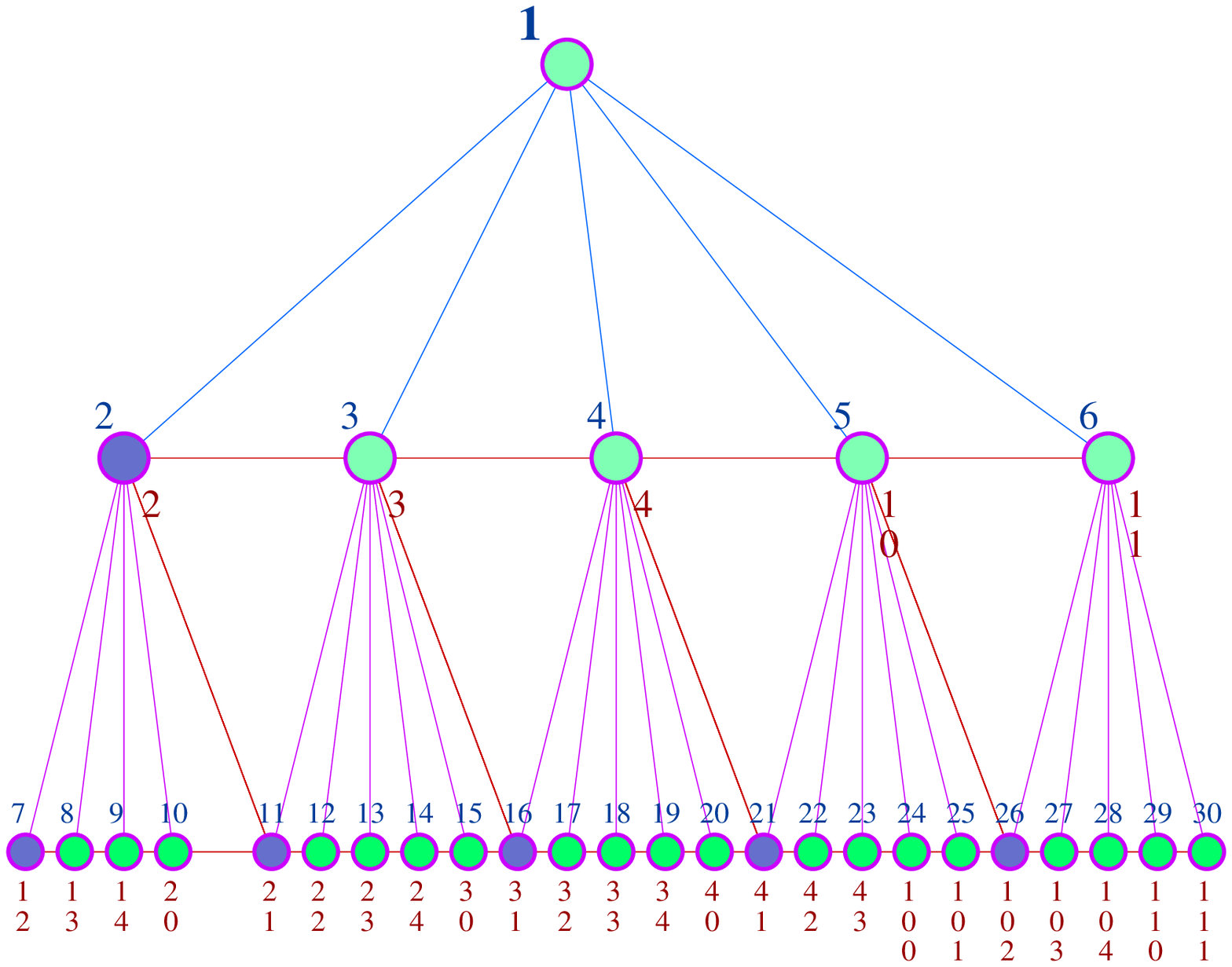}
\hfill}
\ligne{\hfill
\vtop{\leftskip 0pt\parindent 0pt
\vspace{-10pt}
\begin{fig}\label{dualp3L}
\leurre
To left, the leftmost son tree. To right: the dual graph restored from the leftmost son
tree. The dual graph for the tessellation $\{p$$-$$2,4\}$ is obtained by removing the
horizontal red arcs.
\end{fig}
}
\hfill}
}

\vskip 10pt
\vtop{
\ligne{\hfill
\includegraphics[scale=0.33]{tree_9pref.ps}
\includegraphics[scale=0.33]{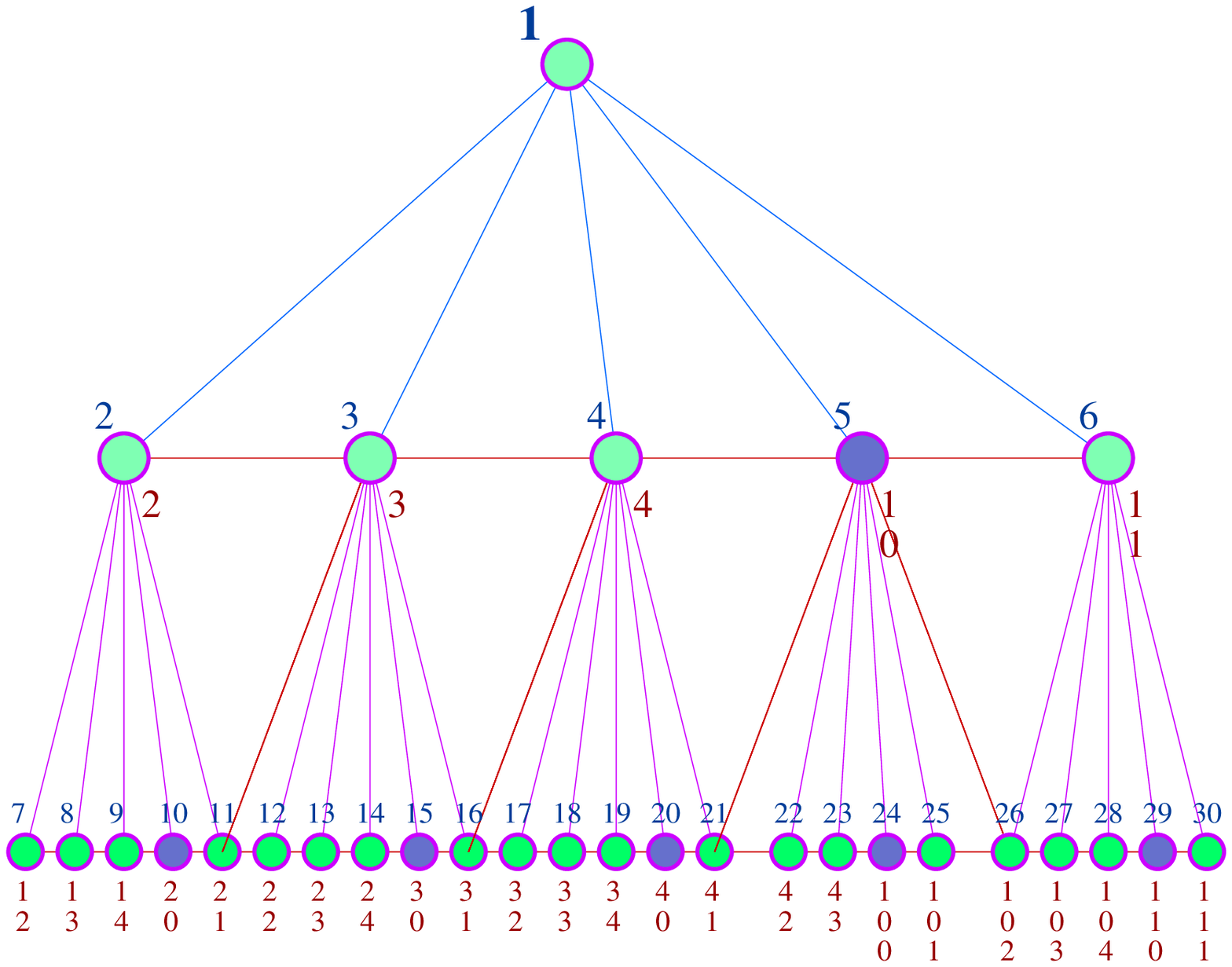}
\hfill}
\ligne{\hfill
\vtop{\leftskip 0pt\parindent 0pt
\vspace{-10pt}
\begin{fig}\label{dualp3P}
\leurre
To left, preferred son tree. To right: the dual graph restored from the preferred son
tree. The dual graph for the tessellation $\{p$$-$$2,4\}$ is obtained by removing the
horizontal red edges.
\end{fig}
}
\hfill}
}
\vskip 5pt
Consider the level~2 in both the leftmost son and the preferred son trees,
$\cal T$ and $\cal P$ respectively. We know the permutation which allows to
pass from one tree to the other one and the shift by one node which it entails
in between the positions on which the permutation operates. Figures~\ref{dualp3L},
\ref{dualp3P} illustrate the transformation of~$\cal T$, $\cal P$ respectively
into the dual graph. Figure~\ref{dualp3LP} displays both restored graphs: it allows
us to compare the processes and to define the transformation from the tree to the
dual graph in the case of the preferred son tree.

On Figure~\ref{dualp3LP}, we can see that the situation is different depending on which
node we consider: the nodes which are the sons of a given node~$\nu$ and which 
occur before the preferred son are applied a rule which is different from the
rule which is applied to the preferred son. 

\vskip 5pt
In $\cal T$, in order to restore the dual graph, we append from each node 
the horizontal arcs which connect the node to its neighbours belonging to the
same level. Outside these arcs, we append an additional one to the right: from the node
to the first son of its right-hand side neighbour. 

In~$\cal P$ we do the same
for the horizontal arcs. The additional arc is appended in a different way.
Consider a node~$\nu$ whose father is~$\mu$, the level of~$\nu$ being~$n$. 
If $\nu$ is a~$W$-node which is not the rightmost son of~$\mu$, its additional arc 
connects it with the rightmost son of~$\nu$$-$1. 
If $\nu$ is a $B$-node, it has two additional arcs, one to the left, connected to the 
rightmost son of~$\nu$$-$1, and one to the right, connected to the leftmost son 
of~$\nu$+1. Hence, if $\nu$ is the rightmost son of~$\mu$, its additional arc
is connected with $\mu$$-$1.
This is illustrated by Figures~\ref{dualp3P} and~\ref{dualp3LP}.
\vskip 5pt
\vtop{
\ligne{\hfill
\includegraphics[scale=0.33]{grtree_9left.ps}
\includegraphics[scale=0.33]{grtree_9pref.ps}
\hfill}
\ligne{\hfill
\vtop{\leftskip 0pt\parindent 0pt
\vspace{-10pt}
\begin{fig}\label{dualp3LP}
\leurre
Comparison of the trees for restoring the dual graph.
To left, the dual graph restored from the leftmost son tree. To right: the dual graph 
restored from the preferred son tree. 
\end{fig}
}
\hfill}
}

We are now in the position to define the coordinates of the neighbours of a node
in both trees, first for the tessellations~$\{p,3\}$.

Consider a node~$\nu$ in $\cal T$. We denote by $f(\nu)$ the father of~$\nu$ and
by $\sigma(n)$ its preferred son. The coordinates are given in 
Table~\ref{voisp3} for $\cal T$ and~$\cal P$. The table also mention the signature 
of the sons of the nodes in each tree.

For the leftmost son tree $\cal T$, we indicate the $W_1$-nodes, the two possible forms
of the $W_2$-nodes and the $B$-node. For the preferred son tree $\cal P$ we indicate the 
three possible $W$-nodes: $W_\beta$ for the leftmost son of its father,
$W_\ell$ for the other $W$-nodes which are to the left of the $B$-son of their father,
$W_r$ for the rightmost son of its father. 

A particular mention is given to the
nodes which are on an extremal branch of the tree: the connections are the same
for both types of tree: the nodes of the leftmost branch receive an arc from the
nodes of the rightmost branch of the previous tree from the previous level. This fixes
the rule for the nodes of the rightmost branch. Table~\ref{voisp4} gives the coordinates
in the case of the tessellations $\{p$$-$$2,4\}$: it is enough to cancel the horizontal
connections. 

\newdimen\largea\largea=25pt
\newdimen\larged\larged=20pt
\newdimen\largec\largec=70pt
\def\lignea #1 #2 #3 #4 #5 #6 #7 #8 #9 {%
\ligne{%
\hbox to\largea{#1\hfill}   
\hbox to\largec{#2\hfill\hbox to \larged{\bf #3\hfill}}   
\hbox to\largec{#4\hfill\hbox to \larged{\bf #5\hfill}}   
\hbox to\largec{#6\hfill\hbox to \larged{\bf #7\hfill}}   
\hbox to\largec{#8\hfill\hbox to \larged{\bf #9\hfill}}   
\hfill}
\vskip 4pt
}
\vtop{
\ligne{\hfill
\vtop{\offinterlineskip\leftskip 0pt\parindent 0pt\hsize=340pt
\begin{tab}\label{voisp3}
\leurre
The neighbours of a node 
for the tessellation $\{p,3\}$.
\end{tab}
\vskip-2pt
\trep
\vskip 8pt
\ligne{\hfill leftmost son tree\hfill}
\vskip 4pt
\lignea {} {$W_1$} {} {$W_2$} {} {$W_2$} {} {$B$} {}
\vskip 4pt
\lignea {0} {$\nu$} {} {$\nu$} {} {$\nu$} {} {$\nu$} {}
\lignea {1} {$f(\nu)$} {} {$f(\nu)$} {} {$f(\nu)$} {} {$f(\nu)$} {} 
\lignea {2} {$\nu$$-$1} {} {$\nu$$-$1} {} {$\nu$$-$1} {} {$f(\nu)$$-$1$^\ast$} {} 
\lignea {3} {$\sigma(\nu)$$-$$p$$+$5} {1} {$\sigma(\nu)$$-$$p$$+$6} {1} 
{$\sigma(\nu)$$-$$p$$+$6} {2} {$\nu$$-$1$^\ast$} {}
\lignea {4} {$\sigma(\nu)$$-$$p$$+$6} {2} {$\sigma(\nu)$$-$$p$$+$7} {2} 
{$\sigma(\nu)$$-$$p$$+$7} {3} {$\sigma(\nu)$$-$$p$$+$6} {2}
\lignea {$p$$-$6} {$\sigma(\nu)$$-$4} {} {$\sigma(\nu)$$-$3} {} 
{$\sigma(\nu)$$-$3} {} {$\sigma(\nu)$$-$4} {}
\lignea {$p$$-$5} {$\sigma(\nu)$$-$3} {} {$\sigma(\nu)$$-$2} {} 
{$\sigma(\nu)$$-$2} {\bii} {$\sigma(\nu)$$-$3} {} 
\lignea {$p$$-$4} {$\sigma(\nu)$$-$2} {\bii} {$\sigma(\nu)$$-$1} {\bii} 
{$\sigma(\nu)$$-$1} {\bun} {$\sigma(\nu)$$-$2} {\bii}
\lignea {$p$$-$3} {$\sigma(\nu)$$-$1} {\bun} {$\sigma(\nu)$} {0} 
{$\sigma(\nu)$} {0} {$\sigma(\nu)$$-$1} {\bun}
\lignea {$p$$-$2} {$\sigma(\nu)$} {0} {$\sigma(\nu)$$+$1} {1} 
{$\sigma(\nu)$$+$1} {1} {$\sigma(\nu)$} {0} 
\lignea {$p$$-$1} {$\sigma(\nu)$$+$1} {} {$\sigma(\nu)$$+$2} {} 
{$\sigma(\nu)$$+$2$^\ast$} {} {$\sigma(\nu)$$+$1} {}
\lignea {$p$} {$\nu$$+$1} {} {$\nu$$+$1} {} {$\nu$$+$1$^\ast$} {} {$\nu$$+$1} {}
\vskip 8pt
}
\hfill}
\ligne{\hfill
\vtop{\offinterlineskip\leftskip 0pt\parindent 0pt\hsize=340pt
\trep
\vskip 8pt
\ligne{\hfill preferred son tree\hfill}
\vskip 4pt
\lignea {} {$W_\beta$} {} {$W_\ell$} {} {$W_r$} {} {$B$} {}
\vskip 4pt
\lignea {0} {$\nu$} {} {$\nu$} {} {$\nu$} {} {$\nu$} {}
\lignea {1} {$f(\nu)$} {} {$f(\nu)$} {} {$f(\nu)$} {} {$f(\nu)$} {} 
\lignea {2} {$f(\nu)$$-$1$^\ast$} {} {$\nu$$-$1} {} {$\nu$$-$1} {} {$\nu$$-$1} {} 
\lignea {3} {$\nu$$-$1$^\ast$} {} {$\sigma(\nu)$$-$$p$$+$5} {} 
{$\sigma(\nu)$$-$$p$$+$6} {2} {$\sigma(\nu)$$-$$p$$+$6} {}
\lignea {4} {$\sigma(\nu)$$-$$p$$+$6} {2} {$\sigma(\nu)$$-$$p$$+$6} {2} 
{$\sigma(\nu)$$-$$p$$+$7} {3} {$\sigma(\nu)$$-$$p$$+$7} {2}
\lignea {$p$$-$6} {$\sigma(\nu)$$-$4} {} {$\sigma(\nu)$$-$4} {} 
{$\sigma(\nu)$$-$3} {} {$\sigma(\nu)$$-$3} {}
\lignea {$p$$-$5} {$\sigma(\nu)$$-$3} {} {$\sigma(\nu)$$-$3} {} 
{$\sigma(\nu)$$-$2} {\bii} {$\sigma(\nu)$$-$2} {} 
\lignea {$p$$-$4} {$\sigma(\nu)$$-$2} {\bii} {$\sigma(\nu)$$-$2} {\bii} 
{$\sigma(\nu)$$-$1} {\bun} {$\sigma(\nu)$$-$1} {\bii}
\lignea {$p$$-$3} {$\sigma(\nu)$$-$1} {\bun} {$\sigma(\nu)$$-$1} {\bun} 
{$\sigma(\nu)$} {0} {$\sigma(\nu)$} {0}
\lignea {$p$$-$2} {$\sigma(\nu)$} {0} {$\sigma(\nu)$} {0} 
{$\sigma(\nu)$$+$1} {1} {$\sigma(\nu)$$+$1} {1} 
\lignea {$p$$-$1} {$\sigma(\nu)$$+$1} {1} {$\sigma(\nu)$$+$1} {1} 
{$\sigma(\nu)$$+$2$^\ast$} {} {$\sigma(\nu)$$+$2} {}
\lignea {$p$} {$\nu$$+$1} {} {$\nu$$+$1} {} {$\nu$$+$1$^\ast$} {} {$\nu$$+$1} {}
\vskip 9pt
\trfn
\vskip 8pt
}
\hfill}
}

\noindent
$^\ast$: The neighbours are different if the node is on an extremal branch. They
belong to another tree: the previous one for~$B$, $W_\beta$ the next one for~$W_2$,
$W_r$.
\vskip 0pt\noindent
For $B$ and $W_\beta$, neighbour~2: $\nu$$-$1; neighbour~3: $\sigma(\nu$$-$$1)$+1.
\vskip 0pt\noindent
For $W_2$ and~$W_r$: neighbour~$p$$-$1: $\nu$$+$1;
neighbour~$p$: $f(\nu)$$+$1. 


\vskip 5pt
\vtop{
\ligne{\hfill
\vtop{\offinterlineskip\leftskip 0pt\parindent 0pt\hsize=340pt
\begin{tab}\label{voisp4}
\leurre
The neighbours of a node 
for the tessellation $\{p$$-$$2,4\}$.
\end{tab}
\vskip -2pt
\trep
\vskip 8pt
\ligne{\hfill leftmost son tree\hfill}
\vskip 4pt
\lignea {} {$W_1$} {} {$W_2$} {} {$W_2$} {} {$B$} {}
\vskip 4pt
\lignea {0} {$\nu$} {} {$\nu$} {} {$\nu$} {} {$\nu$} {}
\lignea {1} {$f(\nu)$} {} {$f(\nu)$} {} {$f(\nu)$} {} {$f(\nu)$} {} 
\lignea {2} {$\sigma(\nu)$$-$$p$$+$5} {1} {$\sigma(\nu)$$-$$p$$+$6} {1} 
{$\sigma(\nu)$$-$$p$$+$6} {2} {$f(\nu)$$-$1$^\ast$} {}
\lignea {3} {$\sigma(\nu)$$-$$p$$+$6} {2} {$\sigma(\nu)$$-$$p$$+$7} {2} 
{$\sigma(\nu)$$-$$p$$+$7} {3} {$\sigma(\nu)$$-$$p$$+$6} {2}
\lignea {$p$$-$6} {$\sigma(\nu)$$-$3} {} {$\sigma(\nu)$$-$2} {} 
{$\sigma(\nu)$$-$3} {\bii} {$\sigma(\nu)$$-$3} {}
\lignea {$p$$-$5} {$\sigma(\nu)$$-$2} {\bii} {$\sigma(\nu)$$-$1} {\bii} 
{$\sigma(\nu)$$-$2} {\bun} {$\sigma(\nu)$$-$2} {\bii} 
\lignea {$p$$-$4} {$\sigma(\nu)$$-$1} {\bun} {$\sigma(\nu)$} {0} 
{$\sigma(\nu)$$-$1} {0} {$\sigma(\nu)$$-$1} {\bun}
\lignea {$p$$-$3} {$\sigma(\nu)$} {0} {$\sigma(\nu)$$+$1} {1} 
{$\sigma(\nu)$$+$1} {1} {$\sigma(\nu)$} {0}
\lignea {$p$$-$2} {$\sigma(\nu)$$+$1} {} {$\sigma(\nu)$$+$2} {} 
{$\sigma(\nu)$$+$2$^\ast$} {} {$\sigma(\nu)$$+$1} {} 
}
\hfill}
\ligne{\hfill
\vtop{\offinterlineskip\leftskip 0pt\parindent 0pt\hsize=340pt
\vskip 8pt
\trep
\vskip 9pt
\ligne{\hfill preferred son tree\hfill}
\vskip 4pt
\lignea {0} {$W_\beta$} {} {$W_\ell$} {} {$W_r$} {} {$B$} {}
\vskip 4pt
\lignea {0} {$\nu$} {} {$\nu$} {} {$\nu$} {} {$\nu$} {}
\lignea {1} {$f(\nu)$} {} {$f(\nu)$} {} {$f(\nu)$} {} {$f(\nu)$} {} 
\lignea {2} {$f(\nu)$$-$1$^\ast$} {} {$\sigma(\nu)$$-$$p$$+$5} {} 
{$\sigma(\nu)$$-$$p$$+$6} {2} {$\sigma(\nu)$$-$$p$$+$6} {}
\lignea {3} {$\sigma(\nu)$$-$$p$$+$6} {2} {$\sigma(\nu)$$-$$p$$+$6} {2} 
{$\sigma(\nu)$$-$$p$$+$7} {3} {$\sigma(\nu)$$-$$p$$+$7} {2}
\lignea {$p$$-$6} {$\sigma(\nu)$$-$3} {} {$\sigma(\nu)$$-$3} {} 
{$\sigma(\nu)$$-$2} {\bii} {$\sigma(\nu)$$-$2} {}
\lignea {$p$$-$5} {$\sigma(\nu)$$-$2} {\bii} {$\sigma(\nu)$$-$2} {\bii} 
{$\sigma(\nu)$$-$1} {\bun} {$\sigma(\nu)$$-$1} {\bii} 
\lignea {$p$$-$4} {$\sigma(\nu)$$-$1} {\bun} {$\sigma(\nu)$$-$1} {\bun} 
{$\sigma(\nu)$} {0} {$\sigma(\nu)$} {0}
\lignea {$p$$-$3} {$\sigma(\nu)$} {0} {$\sigma(\nu)$} {0} 
{$\sigma(\nu)$$+$1} {1} {$\sigma(\nu)$$+$1} {1}
\lignea {$p$$-$2} {$\sigma(\nu)$$+$1} {1} {$\sigma(\nu)$$+$1} {1} 
{$\sigma(\nu)$$+$2$^\ast$} {} {$\sigma(\nu)$$+$2} {} 
\vskip 9pt
\trfn
\vskip 8pt
}
\hfill}
}

\noindent
$^\ast$: The neighbours are different if the node is on an extremal branch. They
belong to another tree: the previous one for~$B$, $W_\beta$ the next one for~$W_2$,
$W_r$.
\vskip 0pt\noindent
For $B$ and $W_\beta$, neighbour~2: $\nu$$-$1. 
\vskip 0pt\noindent
For $W_2$ and~$W_r$: neighbour~$p$$-$2: $\nu$$+$1.
\vskip 10pt
We conclude this section by a visit to the tree which is common to the pentagrid,
the tessellation $\{5,4\}$, and to the heptagrid, the tessellation $\{7,3\}$.
Figure~\ref{dual73LP} displays the trees associated to the leftmost son representation
and the preferred son one.

In the preferred son tree, we can see the following properties:

\begin{thm}\label{pentaheptaP}
When $p=7$ the preferred son tree has the following properties. The tree has two
kinds of nodes, $W$- and $B$-nodes. In $B$-nodes, the preferred son is the leftmost
one, in the $W$-nodes, it is the penultimate. The son signatures are the following ones:
\vskip 5pt
\ligne{\hfill
$\vcenter{
\vtop{\leftskip 0pt\parindent 0pt\hsize=110pt
\ligne{\hfill
\hbox to 40pt{\hfill$B$\hfill}\hbox to 50pt{\hfill$W$\hfill}\hfill}
\ligne{\hfill
\hbox to 40pt{\hfill\bf 0 1\hfill}\hbox to 50pt{\hfill\bf 2 0 1\hfill}
\hfill}
}}$
\hfill{\rm\numlaform}\hskip 20pt}
\newcounter{sighpWBP}
\setcounter{sighpWBP}{0}
\addtocounter{sighpWBP}{\value{laform}}
\end{thm}

\noindent
Proof of Theorem~\ref{pentaheptaP}.
The polynomial we obtain from the decomposition of a sector is this time:
\vskip 5pt
\ligne{\hfill $P(X)= X^2-3X+1$\hfill\numlaform\hskip 20pt}
\newcounter{polyfib}
\setcounter{polyfib}{0}
\addtocounter{polyfib}{\value{laform}}
\vskip 10pt
\vtop{
\ligne{\hfill
\includegraphics[scale=0.4]{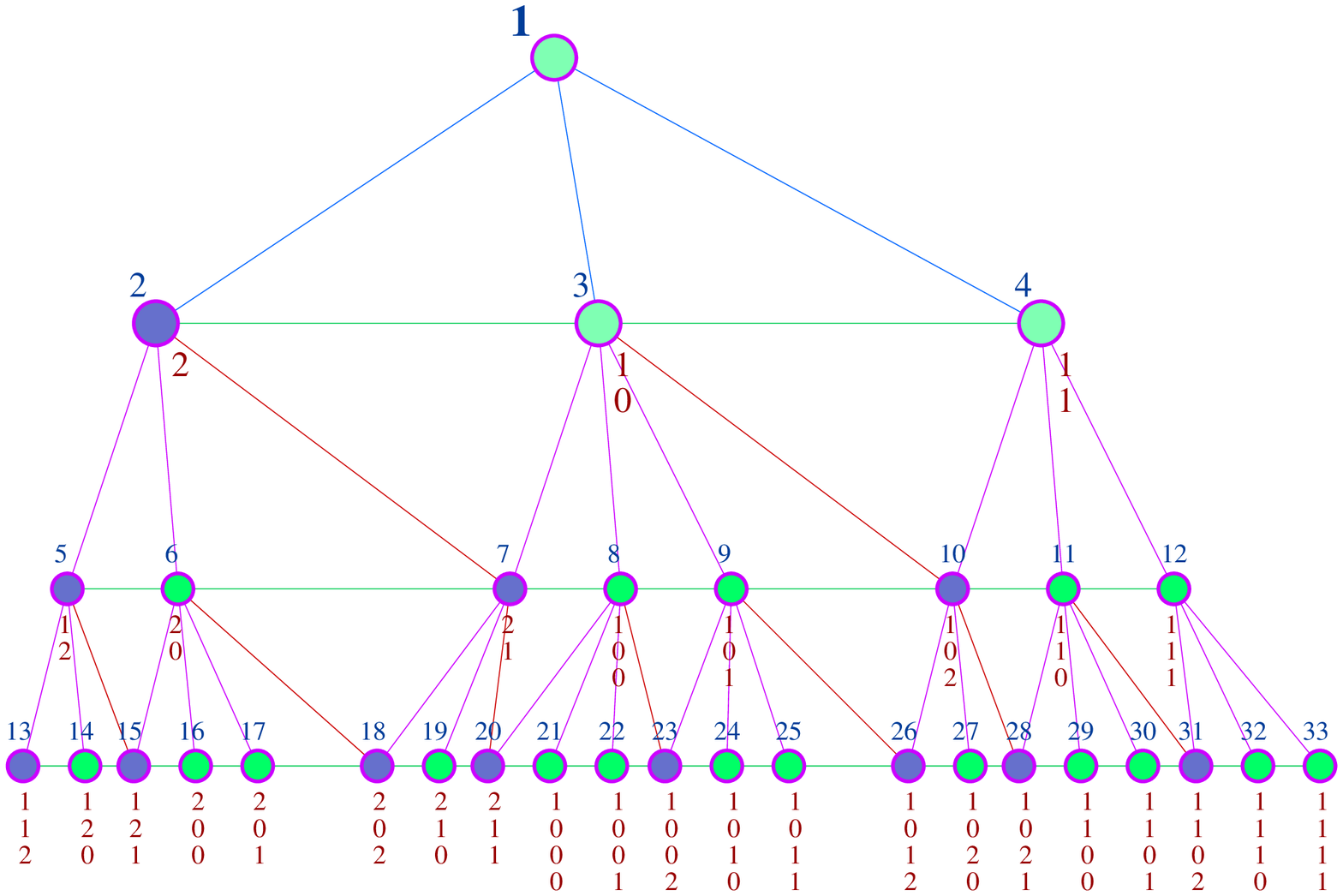}\hfill}
\ligne{\hfill
\includegraphics[scale=0.4]{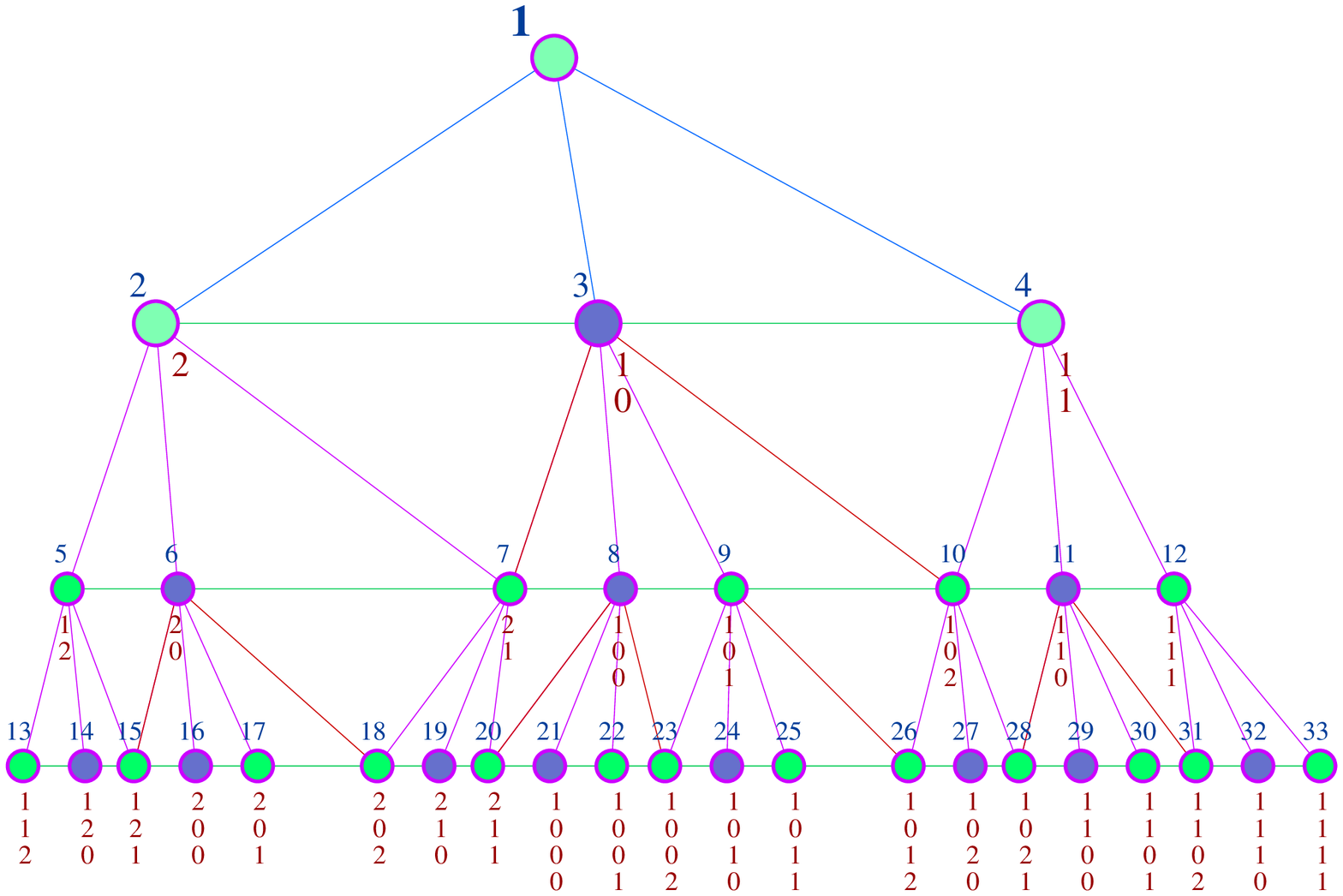}
\hfill}
\ligne{\hfill
\vtop{\leftskip 0pt\parindent 0pt
\vspace{-10pt}
\begin{fig}\label{dual73LP}
\leurre
To left, the leftmost son tree. To right: the dual graph restored from the leftmost son
tree. The dual graph for the tessellation $\{p$$-$$2,4\}$ is obtained by removing the
horizontal red edges.
\end{fig}
}
\hfill}
}
\vskip 5pt

This corresponds to the value $p=7$ in the polynomial obtained from the 
equations~(\therecseq).
From this, it is easy to see that the equation~
(\theunparL) becomes:

\ifnum 1=0 {
\ligne{\hfill
$2u_{n+k}+2u_n+
\displaystyle{\sum\limits_{i=1}^{k-1}u_{n+i}}=u_{n+k+1}+u_{n-1}$
\hfill{\rm\numlaform}\hskip 20pt}
\newcounter{repshort73}
\setcounter{repshort73}{0}
\addtocounter{repshort73}{\value{laform}}

\ligne{\hfill
$2u_{n} +
\displaystyle{\left(\sum\limits_{i=1}^{n-1}u_{i}\right)}
+ 2u_0
=u_{n+1}$\hfill\numlaform\hskip 20pt}
\newcounter{unbis73}
\setcounter{unbis73}{0}
\addtocounter{unbis73}{\value{laform}}
} \fi

\ligne{\hfill
$u_{n+1} = U_n + u_n + 1$
\hfill\numlaform\hskip 20pt}
\newcounter{unparL73}
\setcounter{unparL73}{0}
\addtocounter{unparL73}{\value{laform}}
\vskip 5pt
This tells that the first node on the level~$n$+1 we wind after we crossed the sub-tree 
rooted at the node~2 is the node~$u_{n+1}$. It also confirms that the sequence of
nodes $\{u_n\}_{n\in\mathbb{N}^\ast}$ satisfies the property that $u_{n+1}$ is
a $B$-node which is the preferred son of the node~$u_n$ which is also a $B$-node
for $n>0$, by definition.
Now, formula~(\theunmoinsun) can be rewritten as:
\vskip 5pt
\ligne{\hfill
$2u_{n} +
\displaystyle{\left(\sum\limits_{i=0}^{n-1}u_{i}\right)}
=u_{n+1}-1$\hfill\numlaform\hskip 20pt}
\newcounter{unmoinsun73}
\setcounter{unmoinsun73}{0}
\addtocounter{unmoinsun73}{\value{laform}}
\vskip 5pt
\noindent
telling us that $[u_{n+1}$$-$$1]=\hbox{\bf 21$^n$}$, which also says that the
pattern {\bf 21$^\ast$1} is ruled out in the writing of a coordinate.
From these observations the theorem easily follows. \hfill\boxempty
\vskip 5pt

Turning now to the leftmost son tree, there we can see the following properties:

\begin{thm}\label{pentaheptaL}
When $p=7$ the leftmost son tree has the following properties. The tree has two
kinds of nodes, $W$- and $B$-nodes. In $B$-nodes, the preferred son is the rightmost
one, in the $W$-nodes, it is the penultimate. The son signatures are the following ones:
\vskip 5pt
\ligne{\hfill
$\vcenter{
\vtop{\leftskip 0pt\parindent 0pt\hsize=160pt
\ligne{\hfill
\hbox to 40pt{\hfill$B$\hfill}\hbox to 50pt{\hfill$W_0$\hfill}
\hbox to 50pt{\hfill$W_1$\hfill}\hfill}
\ligne{\hfill
\hbox to 40pt{\hfill\bf 2 0\hfill}\hbox to 50pt{\hfill\bf 1 0 1\hfill}
\hbox to 50pt{\hfill\bf 2 0 1\hfill}
\hfill}
}}$
\hfill{\rm\numlaform}\hskip 20pt}
\newcounter{sighpWBL}
\setcounter{sighpWBL}{0}
\addtocounter{sighpWBL}{\value{laform}}
\vskip 3pt
\noindent
where $W_0$ is a $W$-node whose signature is~{\bf 0} and $W_1$ is
a $W$-node whose signature is~{\bf 1}. Digit~{\bf 2} is always the signature of
a $B$-node.
\end{thm}

\noindent
Proof of Theorem~\ref{pentaheptaL}.
The theorem is a corollary of Theorem~\ref{pentaheptaP} as we can apply to this
tree and recursively to its sub-trees the permutation between the first two nodes in each 
$W$-tree. Accordingly, the son signatures of the first level which
are the first line below become what the second line indicates.
\vskip 3pt
\ligne{\hfill
\bf 2 0 1 -{}- 0 1 -{}- 2 0 1 -{}- 0 1 -{}- 2 0 1 -{}- 2 0 1 -{}- 0 1 -{}- 2 0 1
\hfill} 
\ligne{\hfill
\bf 2 0 -{}- 1 0 1 -{}- 2 0 -{}- 1 0 1 -{}- 2 0 1 -{}- 2 0 -{}- 1 0 1 -{}- 2 0 1
\hfill} 
\vskip 3pt
This distribution is repeated in each sub-tree from one generation to another one.
\hfill\boxempty

\newdimen\llargea\llargea=5pt
\newdimen\llargeb\llargeb=40pt
\def\hhzzv{\hskip 10pt}
\def\hhzzx{\hskip 20pt}
\def\ligneba #1 #2 #3 #4 #5 #6 #7 {%
\ligne{\hfill\small
\hbox to \llargea{#1\hfill} \hhzzv
\hbox to \llargeb{#2} \hhzzv\hbox to \llargeb{#3} \hhzzv
\hbox to \llargeb{#4} \hhzzx\hbox to \llargeb{#5} \hhzzv\hbox to \llargeb{#6}\hhzzv
\hbox to \llargeb{#7} 
\hfill}
}

\vtop{
\begin{tab}\label{vois73LP}
\leurre
The neighbours of a node in the heptagrid, the tessellation~$\{7,3\}$ in both
trees.
\end{tab}
\ligne{\hfill
\vtop{\leftskip 0pt\parindent 0pt\hsize=340pt
\ligne{\hfill\hbox to \llargea{\hfill}\hbox to 160pt{\hfill preferred son tree\hfill}
\hbox to 160pt{\hfill leftmost son tree\hfill}\hfill}
\vskip 3pt
\ligneba {} {$B$\hfill} {$W_b$\hfill} {$W_r$\hfill} 
{$B$\hfill} {$W_0$\hfill} {$W_1$\hfill}
\ligneba {0} {$\nu$\hfill{\bf 0}} {$\nu$\hfill} {$\nu$\hfill} 
{$\nu$\hfill} {$\nu$\hfill{\bf 0}} {$\nu$\hfill}
\ligneba {1} {$f(\nu)$\hfill} {$f(\nu)$\hfill} {$f(\nu)$\hfill} 
{$f(\nu)$\hfill} {$f(\nu)$\hfill} {$f(\nu)$\hfill}
\ligneba {2} {$\nu$$-$1\hfill} {$\nu$$-$1$^\ast$\hfill} {$\nu$$-$1\hfill}
{$f(\nu)$$-$1$^\ast$\hfill} {$\nu$$-$1\hfill} {$\nu$$-$1\hfill} {}
\ligneba {3} {$\sigma(\nu)$$-$1\hfill} {$\sigma(\nu)$$-$1$^\ast$\hfill{\bf 2}} 
{$\sigma(\nu)$$-$1\hfill{\bf 2}} 
{$\nu$$-$1$^\ast$\hfill} {$\sigma(\nu)$$-$1\hfill{\bf 1}} {$\sigma(\nu)$$-$1\hfill{\bf 2}} 
\ligneba {4} {$\sigma(\nu)$\hfill{\bf 0}} {$\sigma(\nu)$\hfill{\bf 0}} 
{$\sigma(\nu)$\hfill{\bf 0}} 
{$\sigma(\nu)$$-$1\hfill{\bf 2}} {$\sigma(\nu)$\hfill{\bf 0}} {$\sigma(\nu)$\hfill{\bf 0}}
\ligneba {5} {$\sigma(\nu)$$+$1\hfill{\bf 1}} {$\sigma(\nu)$$+$1\hfill{\bf 1}} 
{$\sigma(\nu)$$+$1\hfill{\bf 1}}
{$\sigma(\nu)$\hfill{\bf 0}} {$\sigma(\nu)$$+$1\hfill{\bf 1}} 
{$\sigma(\nu)$$+$1\hfill{\bf 1}} 
\ligneba {6} {$\sigma(\nu)$$+$2\hfill} {$\nu$$+$1\hfill} {$\sigma(\nu)$$+$2$^\ast$\hfill} 
{$\sigma(\nu)$$+$1\hfill} {$\sigma(\nu)$$+$2\hfill} {$\sigma(\nu)$$+$2$^\ast$\hfill}
\ligneba {7} {$\nu$$+$1\hfill} {$f(\nu)$$+$1\hfill} {$\nu$$+$1$^\ast$\hfill}
{$\nu$$+$1\hfill} {$\nu$$+$1\hfill} {$\nu$$+$1$^\ast$\hfill}
}
\hfill}
}
\vskip 5pt
As in Table~\ref{voisp3}, with $^\ast$ we indicate that
the neighbours are different if the node is on an extremal branch. They
belong to another tree: the previous one for~$B$, $W_b$ the next one for~$W_1$,
$W_r$.
\vskip 0pt\noindent
For $B$ and $W_b$, neighbour~2: $\nu$$-$1;  neighbour~3: $\sigma(\nu$$-$$1)$+1.
\vskip 0pt\noindent
For $W_1$ and~$W_r$: neighbour~6: $\nu$$+$1; neighbour~7: $f(\nu)$$+$1. 
\vskip 10pt
\vtop{
\begin{tab}\label{vois54LP}
\leurre
The neighbours of a node in the pentagrid, the tessellation~$\{5,4\}$ in both
trees.
\end{tab}
\ligne{\hfill
\vtop{\leftskip 0pt\parindent 0pt\hsize=340pt
\ligne{\hfill\hbox to \llargea{\hfill}\hbox to 160pt{\hfill preferred son tree\hfill}
\hbox to 160pt{\hfill leftmost son tree\hfill}\hfill}
\vskip 3pt
\ligneba {} {$B$\hfill} {$W_b$\hfill} {$W_r$\hfill} 
{$B$\hfill} {$W_0$\hfill} {$W_1$\hfill}
\ligneba {0} {$\nu$\hfill{\bf 0}} {$\nu$\hfill} {$\nu$\hfill} 
{$\nu$\hfill} {$\nu$\hfill{\bf 0}} {$\nu$\hfill}
\ligneba {1} {$f(\nu)$\hfill} {$f(\nu)$\hfill} {$f(\nu)$\hfill} 
{$f(\nu)$\hfill} {$f(\nu)$\hfill} {$f(\nu)$\hfill}
\ligneba {2} {$\sigma(\nu)$$-$1\hfill} {$\sigma(\nu)$$-$1$^\ast$\hfill{\bf 2}} 
{$\sigma(\nu)$$-$1\hfill{\bf 2}} 
{$f(\nu)$$-$1$^\ast$\hfill} {$\sigma(\nu)$$-$1\hfill{\bf 1}} 
{$\sigma(\nu)$$-$1\hfill{\bf 2}} 
\ligneba {3} {$\sigma(\nu)$\hfill{\bf 0}} {$\sigma(\nu)$\hfill{\bf 0}} 
{$\sigma(\nu)$\hfill{\bf 0}} 
{$\sigma(\nu)$$-$1\hfill{\bf 2}} {$\sigma(\nu)$\hfill{\bf 0}} {$\sigma(\nu)$\hfill{\bf 0}}
\ligneba {4} {$\sigma(\nu)$$+$1\hfill{\bf 1}} {$\sigma(\nu)$$+$1\hfill{\bf 1}} 
{$\sigma(\nu)$$+$1\hfill{\bf 1}}
{$\sigma(\nu)$\hfill{\bf 0}} {$\sigma(\nu)$$+$1\hfill{\bf 1}} 
{$\sigma(\nu)$$+$1\hfill{\bf 1}} 
\ligneba {5} {$\sigma(\nu)$$+$2\hfill} {$\nu$$+$1\hfill} {$\sigma(\nu)$$+$2$^\ast$\hfill} 
{$\sigma(\nu)$$+$1\hfill} {$\sigma(\nu)$$+$2\hfill} {$\sigma(\nu)$$+$2$^\ast$\hfill}
}
\hfill}
}
\vskip 10pt
\noindent
$^\ast$ indicates 
that the neighbours are different if the node is on an extremal branch. They
belong to another tree: the previous one for~$B$, $W_b$ the next one for~$W_1$,
$W_r$.
\vskip 0pt\noindent
For $B$ and $W_b$, neighbour~2: $\nu$$-$1;  neighbour~3.
\vskip 0pt\noindent
For $W_1$ and~$W_r$: neighbour~5: $\nu$$+$1. 
\vskip 5pt
In the preferred son tree, the signature of a $B$-node is~always~{\bf 0},
according of the definition of the tree.
The signature of the rightmost son is always~{\bf 1} and the signature of
the leftmost son of a $W$-node is always~{\bf 2}. 

In the leftmost son tree, the situation is not that clear.
\vskip 10pt
   We now turn to the algorithm to compute the branch which leads from the root to
the node~$\nu$ thanks to its coordinate. The algorithms given in the proofs of 
Theorems~\ref{linalgo} and~\ref{linalgoleft} can be simplified in the cases
of the tessellations~$\{p,3\}$ and~$\{p$$-$$2,4\}$. We apply the same strategy
with two auxiliary tables, in order to compute the path from the root to the node
whose coordinates constitute the input of the algorithm. The new algorithms
are not simply deduced from those of Theorems~\ref{linalgo} and~\ref{linalgoleft}
as, for instance the digits {\bf 1} and \bii{} which appear there are different
while here it is the same digit.

In the case of the preferred son and of the leftmost son trees, the algorithm we 
produce here for both the pentagrid and for the heptagrid is different from those 
of Theorems~\ref{linalgo} and~\ref{linalgoleft}, see Algorithms~\ref{a_linalgo7P}
and~\ref{a_linalgoleft}. In those latter algorithms, the notion of slices was clearly
delimited, which is no more the case here. 
Let \hbox{\bf $\pi=~$0$,...,\nu_k$} be a path where {\bf 0} is the root of the tree.
Note that, by definition, in~$\pi$, $\nu_{i+1}$ is a son of~$\nu_i$, with 
\hbox{$i\in[1..k$$-$$1]$} and $\nu_1$ is a son of the root. We say that $\nu_k$ is 
the \textbf{end} of~$\pi$ or that $\pi$ \textbf{leads} from the root
to~$\nu_k$ and we write \hbox{$\pi\models\nu_k$}. Say that $k$+1 is the length
of~$\pi$ also denoted by $\norm\pi\norm$. We say that 
\hbox{\bf $\pi=~$0$,...,\nu_i$} with $i\leq k$ is the \textbf{beginning} of~$\pi$ up 
to~$i$ and we denote it by $\pi_{\vert i}$. We say that $\pi,\mu$ is a 
\textbf{continuation} of $\pi$ if $\mu$ is a son of $\nu$ and only in this case.

\begin{lemm}\label{dist1}
Let $\pi\models\nu$ and $\omega\models\nu$+1, with $\nu$ and $\nu$+1 on the same level
of the tree. We can write \hbox{\bf $\pi=~$0$,...,\nu_k$} and 
\hbox{\bf $\omega =~$0$,...,\mu_k$}.
Then for \hbox{$i\in[1..k]$}, \hbox{$\nu_i\leq\mu_i\leq\nu_i${\rm +1}}.
We shall write \hbox{$\pi\leq\omega\leq\pi${\rm +1}} for that relation.
\end{lemm}

\noindent
Proof of Lemma~\ref{dist1}. The lemma is true for the sons of the root. Assume it 
is true for all nodes up to the level~$n$, that level being included.
Let $\nu$ and~$\nu$+1 be both on the level~$n$+1. Let~$\mu$ be the father of~$\nu$.
If $\nu$+1 is at most the rightmost son of~$\nu$, the property is true. The worst case
is that $\nu$ is the rightmost son of~$\mu$. Then, $\nu$+1 is the leftmost son of $\mu$+1.
Then, by induction, we have a path $\pi\models\mu$ and a path $\omega\models\mu$+1 
satisfying the lemma. Then the paths $\pi,\nu$ and $\omega,\nu$+1 also satisfy the
lemma. \hfill\boxempty

\begin{lemm}\label{sautP}
Consider a node~$\nu$ in the preferred son tree~$\cal P$ and let be $\pi$ with 
$\pi\models\nu$. Then the paths which go to $[\nu]${\bf 0} and to $[\nu]${\bf 1} are 
continuations of~$\pi$. Let $\omega$ be the path leading to~$[\nu]${\bf 2}. Let 
\hbox{$k= \norm\pi\norm$}. Then, \hbox{$\norm\omega\norm = k$$+$$1$}
and \hbox{$\pi\leq\omega_{\vert k}\leq \pi${\rm +1}}.
\end{lemm}

\noindent
Proof of Lemma~\ref{sautP}. The lemma is true for the root and for its sons and also for 
the sons of its sons. From the statement of the lemma, we only have to consider the
case $[\nu]${\bf 2}. Clearly, the signature of~$\nu$ cannot be~{\bf 2}: the pattern
{\bf 22} is ruled out. 

Assume that $[\nu]=[\nu_1]${\bf 0}. Then, the $B$-son of
$\nu_1$ is $[\nu_1]${\bf 0}, its rightmost son $[\nu_1]${\bf 1}. The
sons of $[\nu_1]${\bf 0} are $[\nu_1]${\bf 00} and $[\nu_1]${\bf 01} so that the
sons of $[\nu_1]${\bf 1} are $[\nu_1]${\bf 02}, $[\nu_1]${\bf 10} and $[\nu_1]${\bf 11}.
Hence, $[\nu]${\bf 2} is the leftmost son of $[\nu_1]${\bf 1}. Accordingly,
If $\pi\models\nu_1$, $\pi,[\nu_1]${\bf 1}$,[\nu_1]${\bf 02} satisfy the assumption
of the lemma.

Assume that $[\nu]=[\nu_1]${\bf 1}. Whether $\nu$ is a $B$- or a $W$-node, we know
from~(\thesighpWBP) that its $B$-son is $[\nu_1]${\bf 10} and its rightmost son
is $[\nu_1]${\bf 11} so that \hbox{$[\nu]${\bf 2} = $[\nu_1]${\bf 12}} is the leftmost son
of $[\nu_1]${\bf 2} which is $\nu$+1. From Lemma~\ref{dist1}, if $\omega\models\nu$+1,
writing $\omega=~${\bf 0}$,...,\nu$+1 and $\pi\models\nu$ with
$\pi=~${\bf 0}$,...,\nu$, we have $\pi\leq\omega\leq\pi$+1,
so that $\omega,[\nu_1]${\bf 12} satisfies the conclusion of the
lemma.\hfill\boxempty

   This allows us to justify Algorithm~\ref{a_linalgo7P}.

The idea is to have two paths: $\ell\models [\nu]_{\vert i}$
and $r\models [\nu]_{\vert i}$+1 where $i$ is the current position in a loop
going down one by one from~$k$ to~0, the initialization of $\ell$ and~$r$
being performed in the loop itself. 
As in Algorithms~\ref{a_linalgo} and~\ref{a_linalgoleft},
the paths are defined by digits in $\{1,2\}$ from a $B$-node, in $\{1,2,3\}$ from a 
$W$-node.
An actualization is needed when digits {\bf 0} or {\bf 2} are met. We can check
it on the lemmas: when {\bf 0} is met, we make $r_{\vert i+1}:=\ell_{\vert i+1}$
and $r(i) = \ell(i)$+1, as later we remain in the sub-tree rooted at the node 
reached by $\ell_{\vert i+1}$. When {\bf 2} is met, as the path always goes to right,
we have to consider the sons of the node reached by~$r$, so that this time we 
make $\ell_{\vert i+1} := r_{\vert i+1}$ and again $r(i) = \ell(i)$+1. 
\vskip 5pt
\ligne{\hfill
\vtop{\leftskip 0pt\parindent 0pt\hsize=340pt
\begin{algo}\label{a_linalgo7P}\leurre
Tessellations $\{7,3\}$ and $\{5,4\}$: $p=7$.
Computation of the path from the root to a node in the {\bf preferred son} tree.
\end{algo}
\vskip-2pt
\trep
\vskip 8pt
\ligne{{\sc input}: {\bf a$_k..$ a$_0$}; $\ell$, $r$: tables of size $k$$+$$1$;
$prev := k$; $s_\ell := W$, $s_r := W$;\hfill}
\ligne{\bf \hskip 35pt procedure actualize ($a$, $b$, $i$, $t$) is\hfill}
\ligne{\bf \hskip 35pt begin\hfill}
\ligne{\bf \hskip 50pt for j in $[i..t]$ loop $a(j) := b(j)$; end loop; $t := i$;\hfill}
\ligne{\bf \hskip 35pt end procedure;\hfill}
\ligne{{\bf for $i$ in reverse $[0..k]$}\hfill}
\ligne{\bf loop if a$_i = 0$ {-$\!\,\,$-} $i< k$\hfill}
\ligne{\bf \hskip 35pt then actualize ($r$, $\ell$, $i$$+$$1$, $prev$);\hfill}
\ligne{\bf \hskip 65pt $\ell(i) := 1$; $r(i) := 2$;\hfill}
\ligne{\bf \hskip 65pt if $s_\ell = W$ then $\ell(i) := 2$; $r(i) := 3$; 
end if;\hfill}
\ligne{\bf \hskip 65pt $s_\ell := B$; $s_r := W$; \hfill}
\ligne{\bf \hskip 35pt elsif a$_i = 1$\hfill}
\ligne{\bf \hskip 50pt then if $i=k$ \hfill} 
\ligne{\bf \hskip 85pt then $\ell(k) := 0$; $r(k) := 1$; \hfill}
\ligne{\bf \hskip 85pt else $r(i) := 1$; $\ell(i) := 2$; \hfill}
\ligne{\bf \hskip 110pt if $s_\ell = W$; then $\ell(i) := 3$; end if; \hfill}
\ligne{\bf \hskip 77.5pt end if;\hfill}
\ligne{\bf \hskip 77.5pt $s_\ell := W$; $s_r := W$; \hfill}
\ligne{\bf \hskip 35pt else {-$\!\,\,$-} a$_i = 2$\hfill}
\ligne{\bf \hskip 50pt if $i = k$\hfill}
\ligne{\bf \hskip 57.5pt then $\ell(i) := 1$; $r(i) := 2$; $s_\ell := W$; $s_r := B$;
\hfill}
\ligne{\bf \hskip 57.5pt else {-$\!\,\,$-} $i < k$\hfill}
\ligne{\bf \hskip 72.5pt actualize ($\ell$, $r$, $i$$+$$1$, $prev$);\hfill}
\ligne{\bf \hskip 72.5pt $\ell(i) := 1$; $r_i := 2$; $s_\ell := W$; $s_r := B$;\hfill}
\ligne{\bf \hskip 50pt end if;\hfill}
\ligne{\bf \hskip 26pt end if;\hfill}
\ligne{\bf end loop;\hfill}
\ligne{{\sc output}: $\ell$; \hfill}
\vskip 9pt
\trfn
\vskip 8pt
}
\hfill}

Lemma~\ref{dist1} is also true for the leftmost tree: the same argument holds as
the tree is build by similar rules. Here, the important fact is that black nodes
have one son less than white ones exactly and that each node has one black son exactly.
However, Lemma~\ref{sautP} is no more true as stated for the preferred son tree.
Here we have:

\begin{lemm}\label{sautL}
Consider a node~$\nu$ in the preferred son tree~$\cal P$ and let be $\pi$ with 
$\pi\models\nu$. Then the path which goes to $[\nu]${\bf 0} is a continuation of~$\pi$. 
Let $\omega_\alpha$ be the path leading to~$[\nu]\alpha$ where 
\hbox{\bf $\alpha\in\{$1$,$2$\}$}. Let \hbox{$k= \norm\pi\norm$}. Then, 
\hbox{$\norm\omega_\alpha\norm = k$$+$$1$}
and \hbox{$\pi\leq{\omega_\alpha}_{\vert k}\leq \pi${\rm +1}}.
\end{lemm}

\vskip 5pt
\ligne{\hfill
\vtop{\leftskip 0pt\parindent 0pt\hsize=320pt
\begin{algo}\label{a_linalgo7L}\leurre
Tessellations $\{7,3\}$ and $\{5,4\}$: $p=7$. Computation of the path from the root 
to the node in the {\bf leftmost son} tree.
\end{algo}
\vskip-2pt
\trep
\vskip 8pt
\ligne{{\sc input}: {\bf a$_k..$ a$_0$}; $\ell$, $r$: tables of size $k$$+$$1$;
$prev := k$; $s_\ell := W$, $s_r := W$;\hfill}
\ligne{\bf \hskip 35pt procedure actualize ($a$, $b$, $i$, $t$) is\hfill}
\ligne{\bf \hskip 35pt begin\hfill}
\ligne{\bf \hskip 50pt for j in $[i..t]$ loop $a(j) := b(j)$; end loop; $t := i$;\hfill}
\ligne{\bf \hskip 35pt end procedure;\hfill}
\ligne{{\bf for $i$ in reverse $[0..k]$}\hfill}
\ligne{\bf loop if a$_i = 0$ \hfill} 
\ligne{\bf \hskip 35pt then if $s_\ell = W_0$ or $s_\ell = W_1$\hfill}
\ligne{\bf \hskip 70pt then actualize($r,\ell,i$$+$$1,prev$);\hfill}
\ligne{\bf \hskip 100pt  $r(i) := 3$; $s_r := W_1$;\hfill}
\ligne{\bf \hskip 67pt else $r(i) := 1$; $s_r := B$;\hfill}
\ligne{\bf \hskip 65pt end if;\hfill}
\ligne{\bf \hskip 65pt $\ell(i) := 2$; $s_\ell := W_0$; 
{-$\!\,\,$-} \rm preferred son = 2$^{\rm d}$ son\hfill}
\ligne{\bf \hskip 35pt elsif a$_i = 1$ \hfill}
\ligne{\bf \hskip 50pt then if $s_\ell := W_0$ or $s_\ell = W_1$ \hfill}
\ligne{\bf \hskip 85pt then $\ell(i) := 3$; $r(i) := 1$; $s_\ell = W_1$; $s_r := B$; 
\hfill}
\ligne{\bf \hskip 85pt else actualize($\ell,r,i$$+$$1,prev$);\hfill}
\ligne{\bf \hskip 110pt $\ell(i) := 1$; $r(i) := 2$; $s_\ell := B$; $s_r := W_0$;\hfill} 
\ligne{\bf \hskip 80pt end if;\hfill}
\ligne{\bf \hskip 35pt else {-$\!\,\,$-} a$_i =$~2\rm , $s_\ell = W_0$ or 
$s_\ell = W_1$;\hfill}
\ligne{\bf \hskip 50pt actualize($\ell,r,i$$+$$1,prev$); \hfill}
\ligne{\bf \hskip 50pt $\ell(i) := 1$; $r(i) := 2$; $s_\ell := B$; $s_r := W_0$;\hfill}
\ligne{\bf \hskip 26pt end if;\hfill}
\ligne{\bf end loop;\hfill}
\ligne{{\sc output}: $\ell$; \hfill}
\vskip 9pt
\trfn
\vskip 8pt
}
\hfill}

\noindent
Proof of Lemma~\ref{sautL}. Let $\pi\models\nu$. The fact that the path to~$[\nu]${\bf 0} 
is a continuation of~$\pi$ is a direct corollary of~(\thesighpWBL). 

Consider the case of $[\nu]${\bf 1}. If $\nu$ is a $W_0$ or a $W_1$-node, the path 
to~$[\nu]${\bf 1} is a continuation of~$\pi$. If $\nu$ is a $B$-node, $\nu$+1 is in the 
tree, on the same level of~$\nu$. By Lemma~\ref{dist1}, there is a path~$\omega$
leading to~$\nu$+1 with $\norm\omega\norm=\norm\pi\norm$ and 
\hbox{$\pi\leq\omega\leq\pi$+1}. Now, the path to $\nu${\bf 1} is a continuation
of~$\omega$, so that the lemma is true.

Consider the case of $[\nu]${\bf 2}. Then $\nu$ cannot be a $B$-node. Indeed, if
$\nu$ is a $B$-node, $\nu$+1 is a $W_0$-node. The son signature of~$\nu$+1 is
\hbox{\bf 1 0 1} while the son signature of~$\nu$ is \hbox{\bf 2 0}, see~(\thesighpWBL).
This means that going from $\mu$, the leftmost son of~$\nu$+1, to $\mu$+1 which is the
black son of~$\nu$+1, we go from a coordinate ending in~{\bf 1} to a coordinate
ending in~{\bf 0}. This is possible only if $[\mu]$ has a suffix
of the form~{\bf 21$^\ast$}. Now, this pattern is a suffix of $[\nu]$ as $\nu$+1 ends 
in~{\bf 0}. Accordingly, $[\nu]${\bf 2} is not the coordinate of a node of the tree.
Hence, $\nu$ is a $W$-node. Whatever the status of $\nu$+1, the signature of its 
leftmost son is {\bf 2}. Now as $[\nu]$ is not a prefix of the coordinate of the 
sons of~$\nu$, $[\nu]${\bf 2} is the coordinate of the leftmost son of~$\nu$+1.
The path to $[\nu]${\bf 2} is a continuation of the path to $\nu$+1, so that the
lemma is proved here too.\hfill\boxempty

This allows us to prove the correctness of Algorithm~\ref{a_linalgo7L}. Again,
$\ell$ and~$r$ satisfy \hbox{$\ell\leq r\leq\ell$+1} at the beginning of the body of 
the loop and, at the same moment, we have \hbox{$r(i$$+$$1) = \ell(i$$+$$1)$+1}.  
These conditions are still true at the end of the body of the loop. Note that
the algorithm is a simple translation of the proofs of Lemma~\ref{sautL} and that
the actualization is needed when $\ell$ continues its previous value or when
the continuation requires to take~$r$. The actualization is here not symmetric,
contrarily to what can be seen in Algorithm~\ref{a_linalgo7P}. Indeed, it was proved 
in~\cite{mmbook2} that the path from the root of the tree to a node is the leftmost
one in the leftmost son tree. This means that if $\pi\models\nu$ and $\omega\models\nu$
with $\pi$ in $\cal P$ and $\omega$ in $\cal T$, then if $\mu\in\omega$ and $\nu\in\pi$
are on the same level, then $\nu\leq\mu$.
 
\section{Conclusion}

These tools offer the possibility to define convenient coordinates for the
study of the tessellations $\{p,3\}$ and $\{p$$-$$2,4\}$. In the case of the pentagrid
and of the heptagrid, I used another system based on Fibonacci numbers. The
connections between the Fibonacci coordinates and those indicated in this paper,
namely in the last part of Section~\ref{res}. Note that the greatest root of
the polynomial~(\thepolyfib) is the square of the golden ratio from which the Fibonacci
sequence can be obtained.

\end{document}